\documentclass[prb,aps,twocolumn]{revtex4}

\usepackage{graphicx}
\usepackage{amsmath}
\usepackage{epstopdf}
\usepackage{amsbsy}
\usepackage{appendix}
\usepackage{ulem}
\usepackage{color}

\newcommand{\beq}{\begin{equation}}
\newcommand{\eeq}{\end{equation}}
\newcommand{\bea}{\begin{eqnarray}}
\newcommand{\eea}{\end{eqnarray}}
\newcommand{\e}{\varepsilon}
\newcommand{\bk}{{\vec k}}

\newcommand{\bq}{{\vec q}}

\begin{document}

\title{Nematicity and Superconductivity: Competition 
vs. Cooperation}
\author{Xiao Chen$^1$, S. Maiti$^{1,2}$, R. M. Fernandes$^3$, P. J. Hirschfeld$^1$}
\affiliation{ $^1$Department of Physics,  U. Florida, Gainesville FL 32611 USA}
\affiliation {$^2$Department of Physics, Concordia University, Montreal, QC H4B1R6, Canada}
\affiliation{$^3$School of Physics and Astronomy, University of Minnesota, Minneapolis 55455, MN}

\begin{abstract}
Electronic nematic behavior has been identified and studied in iron-based superconductors for some time, particularly in the well-known BaFe$_2$As$_2$ system, where it is well-known to compete with superconductivity.   On the other hand, it has been shown recently that FeSe displays a negligible effect of nematicity on superconductivity near the superconducting transition, and actual cooperation between the two orders when the system is doped with S. Recently it has also been proposed that LiFeAs undergoes a nematic transition in the superconducting state itself. Generally, we expect  superconductivity to be anisotropic when it coexists with nematic order, but it is not clear under what circumstances the two orders compete or cooperate, nor how the anisotropy of the superconducting state correlates with that in the nematic state. To address this, we study a simple mean field model of a $d$-wave Pomeranchuk instability together with a mixed $s,d$ pairing interaction, and identify when nematicity is enhanced or suppressed by superconductivity. We show that the competition or cooperation depends significantly on the  distortion of the Fermi surface due to nematicity relative to the anisotropy of the superconducting gap function. Further, we discuss the implications of our results for FeSe and LiFeAs.
\end{abstract}
\maketitle{}

\section{Introduction}

Electronic nematic order in iron-based superconducting (FeSC) materials has been the subject of considerable interest for several years now, after being established by several key STM\cite{Chuang_Davis2010}, thermodynamic\cite{Boehmer_review_CRPhysique} and transport experiments\cite{Chu_Fisher_Science2012}.   In general, nematic instabilities in the FeSC have been
discussed in terms of a competition between fluctuations of  structural, orbital,
and spin degrees of freedom\cite{Fernandes_NatPhys_2014,HuXu_review2016}. Particularly in the Fe pnictides, however, the proximity of the structural and magnetic transitions, as well as the observed scaling of the magnetic and lattice fluctuations in these systems\cite{Fernandes_SchmalianPRL2013}, led to the general idea that the 
nematic phase preceding magnetic ordering is driven by spin fluctuations. 

On the other hand, the origin of the nematic order in iron chalcogenides like FeSe is still controversial.     At first glance,
the absence of long-range magnetic order in the ambient pressure phase diagram of FeSe suggests
 that the spin nematic paradigm\cite{Christensen16} might not be
appropriate, and that orbital fluctuations might play a more leading role\cite{bohmer13,Kontani15,Fanfarillo2018}.
On the other hand, the confirmation of a long-range magnetic state under a modest
pressure\cite{Bendele12,Terashima15} has lent support to other proposals that suggest
that the  ambient pressure phase may be a quantum paramagnet \cite{FWang15,Glasbrenner15}
or possibly a state with long-range order of ``hidden'' magnetic quadrupolar type\cite{Yu15,Nevidomskyy16}.
The tiny Fermi surfaces in this system may also be important to prevent long-range magnetic ordering\cite{Chubukov_smallFS15,Chubukov2016PRX,Glasbrenner15}.

Beyond  addressing the origin of the nematic state, it is interesting to ask what the influence of nematicity  is on the superconductivity that evolves out of it.
Fernandes and Millis\cite{Fernandes_Millis_PRL2013} studied the problem of the coupling of the  nematic, $s$- and $d$-wave order parameters, and found  several phase diagrams illustrating the transition from $s$-wave to $d$-wave pairing with varying coupling to nematicity, according to whether the nematic order was condensed or fluctuating. The phenomenological study addressed the scenario when the nematic and superconducting transition temperatures ($T_{n},T_c$) are close to each other. {Here we discuss results based on a simple microscopic model that 
describes the basic physics over the entire range of the bare ratio $T_n/T_c$.}


The question of whether the two orders compete or cooperate has been raised with new urgency recently by several key experimental results {on the Fe-based systems FeSe and LiFeAs}. The first is an electron irradiation experiment by the Prozorov group\cite{Teknowijoyo16} that showed that disorder, surprisingly, enhances $T_c$ slightly, in contrast to similar experiments in Fe pnictides. These authors discussed various possible explanations for their observation, including the possibility of a competition of nematic and superconducting order, that might allow a $T_c$ enhancement if the nematic order were to be suppressed   more rapidly by disorder, by analogy to superconductivity competing with density wave order. This scenario was explored by Mishra and Hirschfeld\cite{Mishra_NJP2016}, who found that it could occur in a simple model where nematic order is driven by a $d$-wave Pomeranchuk instability if there were strong {\it competition} between nematic and superconducting order.  The degree of competition was found to depend strongly on the orientation of the nematic director relative to the superconducting anisotropy.
 
The second set of experimental results come from the Karlsruhe group, who reported a surprising lack of coupling between the orthorhombic $a,b$ axis lattice constant splitting  (indicative of nematicity) and superconductivity in FeSe; in contrast to Ba-122, where the splitting was suppressed below $T_c$, indicating competition\cite{Nandi2010}. In FeSe, there was only a minor effect on $\delta = a-b$ at $T_c$\cite{Meingast_no_coupling_FeSe}.  Even more surprising was a recent new result on FeSe doped with sulfur reported  by the same group in Ref. ~\onlinecite{Meingast_S-dopedFeSe}.  Although with increased S doping, $T_c$ is known to increase while $T_{n}$ decreases, suggesting competition, $\delta$ was found to {\it increase} as $T$ was lowered below $T_c$, indicating a {\it cooperative} effect of superconductivity and nematicity in these samples.  Phase cooperation  is not particularly  common when superconductivity is involved, but occurs in some other contexts\cite{Paolasini2002}.

Finally,  an angle-resolved photoemission experiment on the {\it tetragonal} compound LiFeAs indicated recently for the first time that nematicity could occur below $T_c$, with the measurement of a strongly $C_2$-symmetric gap function\cite{Kushnirenko2018}.  While data at only one temperature below $T_c$ were reported, these authors speculated that the nematicity might appear spontaneously at $T_c$, and gave a Ginzburg-Landau (GL) argument how this could occur.  The theoretical proposal of such nematic order induced by superconductivity was also discussed by Livanas et al\cite{Livanas2015},  and the role of orbital nematic fluctuations on superconductivity was explored by Yamase and Zeyher \cite{Yamase2013}.  The influence of a leading Pomeranchuk instability on the superconducting $T_c$ in the Hubbard model was discussed in terms of possible cooperation or competition by Kitatani et al\cite{Kitatani2017}.

Evidently a wide range of behavior is possible, and we propose to investigate the phase diagram of a simple model that allows for both competition and cooperation. Such a model, {while simple}, can form the foundation for more challenging studies which could address the role of disorder, orbitals and other electronic correlations, and incorporate the induced nematic distortion of the Fermi surface (FS) into the pairing interaction itself. The basic question we address here in the simpler situation is, ``what aspects of nematic order and superconductivity
influence whether these two phenomena cooperate or compete with one another?"

 To minimize the number of parameters, we work primarily with a single band system. The minimal ingredients needed to study the above question are tendencies towards nematic, $s$- and $d$- wave superconducting orders. We emphasize that we do not address the mechanism that can lead to these tendencies but simply assume an effective theory where the electronic correlations have already resulted in the above mentioned channels being attractive. A more complete  treatment would enable one to model the system with respect to microscopic variables, e.g. the carrier concentration (see, e.g. Ref. \onlinecite{Referee2}), but our current choice expressing results in terms of effective $s$- and $d$-wave interactions is  more transparent and simpler.  The evolution of $s$ and $d$ spin fluctuations with doping has been discussed in Ref. \onlinecite{Maiti2011}.

Our results for the one-band model, in the absence of disorder, can be summarized as follows:
\begin{itemize}
\item It is possible to have a superconducting order emerge out of the nematic order, however, the two orders cooperate only if the anisotropy of the superconducting order parameter is such that the direction of the gap maximum aligns with the elongation of the Fermi surface distortion arising from the nematic order, as shown in Fig  \ref{fig:sketch1}.
\begin{figure}[htp]
$\begin{array}{c}
\includegraphics[width=0.25\columnwidth]{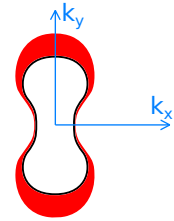}
\end{array}$
 \caption{
\label{fig:sketch1} A sketch of a FS (black curve) and the superconducting gap around it (width of the red region), with the FS elongation due to nematicity and the gap maximum in the same direction.}
\end{figure}
\item The degree of cooperation is generally quite small, and is controlled by the high-energy sector of the electronic spectrum.  While this is true for the one band model, the question remains open for multi-component systems in general.
\item The cooperative effect can exist even when the nematic order emerges out of the superconducting order. However, we need strong competition of $s$- and $d$-wave orders to see this effect.
\end{itemize}
We also analyse a multiband scenario with one hole and two electron pockets,   demonstrating a cooperative effect for a similar condition of alignment of the gap-maximum and FS distortion, as in the one band case. 

The rest of the text is organized as follows. In section \ref{sec:GL} we review the Ginzburg-Landau approach to emphasize how a cooperative behavior may emerge. In Sec. \ref{sec:model} we describe our band model with the appropriate correlations. In Sec. \ref{sec:A} we characterize the nematic state, and then address the onset of superconductivity in the nematic state ($T_n>T_c$). We solve the self consistent equations involving both order parameters and check the free energy to ensure the stability of the solution. In Sec. \ref{sec:B} we discuss the case with $T_n<T_c$. In Sec. \ref{sec:exp} we put our results in the context of current experiments and other works. In Sec. \ref{sec:Conclusion} we present our summary and an outlook for future works.

\section{Origin of cooperative effect}\label{sec:GL}
\begin{figure}[htp]
$\begin{array}{c}
\includegraphics[width=0.9\columnwidth]{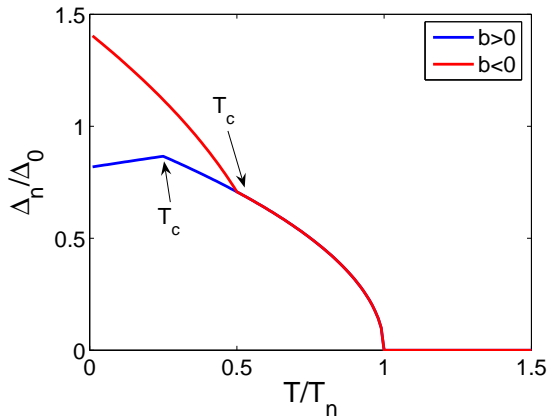}
\end{array}$\caption{
\label{fig:0} Enhancement or suppression of the nematic order parameter with T as superconductivity sets in. Here $c_{s,n}=1/\Delta_0^3$, $|b|=0.5/\Delta_0^2$, $a_{s,n}=a_{s,n}^0(T/T_{s,n}-1)$ and $T_n=\Delta_0$, $T_s=0.4\Delta_0$, $a_{s,n}^0=1/\Delta_0$ and $\Delta_0=\Delta_n(T=0)$.}
\end{figure}

The fact that nematicity and superconductivity can cooperate can be seen at the level of a GL analysis. In the GL regime, it has been known that nematicity induces new coupling between various superconducting orders\cite{Fernandes_Millis_PRL2013}. To see which parameter in the GL theory controls competition vs. cooperation, let us look first at a specific model where the only attractive superconducting channel is isotropic $s$-wave.  The free energy describing the coupling of an $s$-wave SC order parameter $\Delta_s$ to a nematic order parameter $\Delta_n$:
\bea\label{eq:1}
\mathcal{F}&=&\frac{a_s}{2}\Delta_s^2 + \frac{a_n}{2}\Delta_n^2+\frac{b}{2}\Delta_s^2\Delta_n^2 + \frac{c_s}{4}\Delta_s^4 + \frac{c_n}{4}\Delta_n^4.
\eea
Note that Eq. (\ref{eq:1}) is valid regardless of whether one is in the disordered, nematic, or superconducting phases, as long as the temperature is not too far from the (nearly degenerate) transition temperatures.
For stability of the individual and coexistence phases, we will impose $c_{n,s}>0$ and $c_nc_s>b^2$. {If nematic order sets in first}, $a_n<0$ and $a_s>0$, leading to $\Delta_s=0$ and $\Delta_n^2=|a_n|/c_n$. The  subsequent $s$-wave transition is shifted from $a_s\rightarrow0$ to $a_s=-b|a_n|/c_n$. The respective strengths of the order parameter are given by:
\bea\label{eq:2}
\Delta_s^2&=&\frac{1}{c_sc_n-b^2}\left(-b|a_n|-c_na_s\right),\\
\Delta_n^2&=&\frac{1}{c_sc_n-b^2}\left(ba_s+c_s|a_n|\right).
\eea
Observe that when $b>0$, the presence of one order suppresses the onset of the other order as it costs the system energy to accommodate both. {On the other hand}, when $b<0$ the system prefers to have both orders. Figure \ref{fig:0} shows these two cases {schematically} (exaggerated to demonstrate the effect), where we see that the competitive case also leads to suppression of $T_c$ and the cooperative case enhances $T_c$. 
Thus, within a GL description, the switch is the phenomenological coefficient $b$ which can change cooperation to competition with a change of sign. In simple microscopic models,  the parameter $b>0$~\cite{FernandesMaiti2013}, however.

A possible mechanism that can reverse the sign of $b$ is as follows (see also Ref. \onlinecite{Fernandes_Millis_PRL2013}). Consider again the above system with a competing $d$-wave state. We impose that without any coupling between the orders, $T_n>T_c^{(s)}>T^{(d)}_c$. The free energy then acquires the form\cite{Fernandes_Millis_PRL2013}:
\bea\label{eq:4}
\mathcal{F}&=&\frac{a_s}{2}\Delta_s^2+\frac{a_d}{2}\Delta_d^2 + \frac{a_n}{2}\Delta_n^2+\frac{b}{2}\Delta_s^2\Delta_n^2 +\frac{b_d}{2}\Delta_d^2\Delta_n^2 \nonumber\\
&&+\lambda\Delta_s\Delta_d\Delta_n\cos\Theta + \frac12(\alpha+\beta\cos2\Theta)\Delta_s^2\Delta_d^2 \nonumber\\
&&+\frac{c_s}{4}\Delta_s^4+\frac{c_d}{4}\Delta_d^4+ \frac{c_n}{4}\Delta_n^4.
\eea
Here $\Theta$ is the relative phase between $\Delta_s$ and $\Delta_d$. Without $\Delta_n$, we expect $\Theta=0$ or $\pi/2$; furthermore there are $\Theta=0$ solutions with mixed $s$ and $d$ symmetry~\cite{Musaelian1996,Livanas2015}.   Because $T_n>T_c^{(s)}>T^{(d)}_c$, we can choose to focus around the regime where $T\sim T_c^{(s)}$ and $\langle\Delta_d\rangle=0$. In this regime, we can integrate out the effect of $d-$wave fluctuations (by ignoring the quartic term and integrating the resulting Gaussian action with respect to real and imaginary parts of $\Delta_d$). This results in a free energy of the form in Eq. (\ref{eq:1}) but with a modified $b$ coefficient:
\bea\label{eq:44}
\mathcal{F}&=&\frac{a_s}{2}\Delta_s^2+ \frac{a_n}{2}\Delta_n^2+\left(\frac{b}{2}-\frac{\lambda^2}{2\tilde a_d}\right)\Delta_s^2\Delta_n^2\nonumber\\
&&+\frac{c_s}{4}\Delta_s^4+ \frac{c_n}{4}\Delta_n^4.
\eea
Thus we see that $b_{\rm eff}=b-\lambda^2/\tilde a_d$, where $\tilde a_d=a_d+(\alpha+\beta)\Delta_s^2 + b_d\Delta_n^2$. If the $d-$wave component is not a competing superconducting instability, {($T_c^{(d)}\ll T_c^{(s)}$), then $1/\tilde a_d$ is small and $b_{\rm eff}\sim b>0$.} However, if the $d-$wave component is a closely competing subleading instability,  {($T_c^{(d)}\lesssim T_c^{(s)}$)}, then $\tilde a_d\rightarrow 0+$ and $b_{\rm eff}$  eventually becomes negative. Thus the proximity of a competing fluctuating state that couples to the nematic and superconducting order parameter effectively turns the competition between nematicity and superconductivity into cooperation. Within the context of the GL theory, the circumstances under which $b$ changes sign were discussed in more detail recently by Labat et al.\cite{Labat2020}.

The limitation of the above analysis is that it requires all the transition temperatures to be close to each other and does not provide details about the band-topology and/or the gap structure necessary to see this effect in a real system. The experimental support for the cooperation phenomenon comes from Ref. \onlinecite{Meingast_S-dopedFeSe}, but the situation there was far from the GL regime. It is thus desirable to go beyond GL analysis and ask if the effect still exists and if it is due to the same reason (competition of $s$- and $d-$ wave orders).

\section{Model description}\label{sec:model}
Consider a single band with dispersion $\xi_{\bk}$, chemical potential $\mu$ and interaction terms that can lead to superconductivity and nematicity. The nematic state here is modeled as a $d$-wave Pomeranchuk state (this is the simplest model that can capture the effect of rotational symmetry breaking on the superconducting state). For the superconducting part, other than the usual $s$-wave interaction, we include a $d$-wave component. 
Thus, a toy model to study the interplay between nematicity and superconductivity can be written down as the following effective Hamiltonian:
\bea\label{eq:starting H}
H&=&\sum_{\bk s}(\xi_{\bk}-\mu)c_{\bk s}^{\dagger}c_{\bk s}+H_{\rm int}^{\rm SC} + H_{\rm int}^{\rm nem},\\
H_{\rm int}^{\rm nem}&=&-\frac14\sum_{\bk\bk'ss'}V_{\bk\bk'}^{nem}
c_{\bk s}^{\dagger}c_{\bk s}c_{\bk's'}^{\dagger}c_{\bk's'},\nonumber\\
H_{\rm int}^{\rm SC}&=&\frac14\sum_{\bk\bk'ss'tt'}V_{\bk\bk'}^{sc}c_{\bk s}^{\dagger}c_{-\bk s'}^{\dagger}c_{-\bk't}c_{\bk't'}\sigma^y_{ss'}\sigma^y_{tt'},\nonumber
\eea
where 
\bea\label{eq:start}
V^{\rm nem}_{\bk\bk'}&=&V^{\rm nem}f_{\bk}f_{\bk'},\nonumber\\
V^{sc}_{\bk\bk'}&=&V^s + V^df_{\bk}f_{\bk'}.
\eea
Here $f_{\bk}=\sqrt{2}\cos2\theta_{\bk}$ and superconductivity is assumed to exist only in the spin singlet channel. In this model, the renormalized interactions $V^{\rm nem}>0$ leads to attraction in the nematic channel and $V^{s,d}<0$ leads to attraction in the superconducting channel.

We now divide our analysis into two scenarios: (a)  superconductivity condenses inside the nematic state, $T_n>T_c$; and (b)  nematicity kicks in inside the superconducting state, $T_n<T_c$. When we study scenario (a), where the nematic order sets in first, we note that we can no longer use the C$_4$ symmetric form $V^{sc}_{\bk\bk'}=V^s + V^df_{\bk}f_{\bk'}$. We expect the superconducting channel to experience a feed-back from the symmetry broken nematic state. A full self-consistent treatment of this effect is outside the scope of this work, but to zeroth order, we can expect the feedback to be modeled by  
\bea\label{eq:start2}
V^{sc}_{\bk\bk'}&=&V^{sc}\mathcal{Y}_{\bk}\mathcal{Y}_{\bk'},
\eea
where $\mathcal{Y}_{\bk}=(1+r\cos2\theta_{\bk})/\sqrt{1+r^2/2}$, normalized over the Fermi surface. In this form, we capture the mixing of the $d-$wave component with the $s$-wave one induced due to nematicity. Thus the pairing anisotropy coefficient $r\propto \Phi_0\sim V^{\rm nem}$ must be zero if the strength of the nematic order parameter $\Phi_0$ is zero, and the pairing interaction reduces to a pure $s$-wave; but in the nematic phase it is generally nonzero and can be  of either sign. In  principle, $r$ is controlled by temperature, $V^{\rm nem}$, and other details of the electronic structure. 

For scenario (b), however, the C$_4$ symmetric form is valid prior to the onset of nematicity. When nematicity sets in, we assume that it is weak enough to not alter the pairing interaction significantly so that Eq. (\ref{eq:start}) can still be used.

\section{Scenario 1: $T_n>T_c$}\label{sec:A}
We first start by looking at the nematic state that sets in before the superconducting state.
\subsection{The Nematic state}
The mean-field assumption leads to a term in the Hamiltonian,

\bea
H^{\rm nem}_{\rm MF}&=&\sum_{\bk s}
\Phi_0f_{\bk}c^{\dagger}_{\bk s}c_{\bk s},\label{eq:MF1}\\
\Phi_0&=& -V^{\rm nem}\sum_{\bk}\langle f_{\bk} c^\dagger_{\bk}c_{\bk}\rangle\nonumber\\
&=&V^{\rm nem}\sum_{\bk}\frac{f_{\bk}}{2}\left[\tanh\left(\frac{\tilde{\e}_{\bk}}{2T}\right)-1\right],\label{eq:MF1a}
\eea
where $\e_{\bk}=\xi_{\bk}-\mu$, $\tilde{\e}_{\bk}=\e_{\bk}+\Phi_0f_{\bk}$. Near $T_n$, $\Phi_0\rightarrow 0$ and we get
\bea\label{eq:MF2}
\Phi_0&=& \frac{V^{\rm nem}\Phi_0}{2}\sum_{\bk}\frac{f^2_{\bk}}{2T_n}{\rm sech}^2\left(\frac{\e_{\bk}}{2T_n}\right).
\eea
Notice that unlike typical weak-coupling instabilities, the RHS of Eq. (\ref{eq:MF2}) does not have any essential singularity. This means that $V^{\rm nem}$ needs to exceed a threshold for the nematic instability to occur. This is an artifact of our model (as opposed to a model where an RG like enhancement can lead to a similar instability\cite{Chubukov2016PRX}). A more systematic treatment would be where such an instability can be driven by growth of $V^{\rm nem}$ under an RG flow/RPA renormalization. Such a treatment is not the subject of our study. If we treat our band as parabolic ($\e_{\bk}=k^2/2m-\mu$), then $T_n=-\mu/[\ln{(\lambda_{\rm nem}-1)}]$, where $\lambda_{\rm nem}=\nu_{0}V^{\rm nem}$ and $\nu_{0}=m/2\pi$ is the density of states at the Fermi surface. It is clear from Eq. (\ref{eq:MF2}) that to obtain a finite $T_n$ we need $\lambda_{\rm nem}>1$. In the interval $1<\lambda_{\rm nem}<2$, we have $0<T_n<\infty$. Thus within our model $1<\lambda_{\rm nem}<2$.  

The free energy of the nematic state relative to the normal state is given by (see derivation in Appendix:\ref{app:b})
\bea\label{eq:FENem}
\Delta \mathcal{F} &=&\mathcal{F}_{\rm nem}-\mathcal{F}_{\rm normal}\nonumber\\
&=&\sum_{\bk}\left\{-T\ln\left[\frac{\cosh^2(\tilde\e_{\bk}/2T)}{\cosh^2(\e_{\bk}/2T)}\right] +\Phi_0f_{\bk}\right\}\nonumber\\
&&~~~~~~~~~~~~~~~ 
+ \sum_{\bk}\frac{\Phi_0f_{\bk}}{2}\left[\tanh\frac{\tilde\e_{\bk}}{2T}-1\right]\nonumber\\
&=&\sum_{\bk}\left\{-T\ln\left[\frac{\cosh^2(\tilde\e_{\bk}/2T)}{\cosh^2(\e_{\bk}/2T)}\right] +\Phi_0f_{\bk}\right\} \nonumber\\
&&~~~~~~~~~~~~~~~ + \frac{\Phi_0^2}{V^{\rm nem}}.
\eea
To arrive at the last line we have used Eq. (\ref{eq:MF1a}). 

\subsection{Coexistence of nematicity and superconductivity}
We now include the effect of the term $H_{\rm int}^{\rm SC}$. Upon a mean-field approximation, we arrive at the following equations:
\bea
H^{\rm SC}_{\rm MF}&=&\frac12\sum_{\bk s}
\Delta_0\mathcal{Y}_{\bk}~s c^{\dagger}_{\bk s}c^{\dagger}_{-\bk \bar s} ~+~{\rm h.c.},\label{eq:MF4}\\
\Delta_0&=& -V^{sc}\sum_{\bk}\mathcal{Y}_{\bk}\langle c_{-\bk\uparrow}c_{\bk\downarrow}\rangle\nonumber\\
&=&-V^{sc}\Delta_0\sum_{\bk}\frac{\mathcal{Y}^2_{\bk}}{2E_{\bk}}\tanh\frac{E_{\bk}}{2T},\label{eq:MF4a}
\eea
where $E_{\bk}=\sqrt{\bar\e^2_{\bk}+\Delta_0^2\mathcal{Y}^2_{\bk}}$. The information about the nematic order is in $\bar\e_{\bk}=\e_{\bk}+\bar\Phi_0f_{\bk}$. To have a superconducting solution, we need $V^{sc}<0$. The presence of $\Delta_0$ also changes Eq. (\ref{eq:MF1a}) to
\bea\label{eq:MF5}
\bar\Phi_0&=& -V^{\rm nem}\sum_{\bk}\langle f_{\bk} c^\dagger_{\bk}c_{\bk}\rangle\nonumber\\
&=&V^{\rm nem}\sum_{\bk}\frac{f_{\bk}}{4}\left[\left\{1+\frac{\bar\e_{\bk}}{E_{\bk}}\right\}\tanh\left(\frac{E_{\bk}}{2T}\right)\right.\nonumber\\
&&~~~~~~~~~~~~~~~\left. +\left\{1-\frac{\bar\e_{\bk}}{E_{\bk}}\right\}\tanh\left(\frac{-E_{\bk}}{2T}\right)-2\right],\nonumber\\
&=&V^{\rm nem}\sum_{\bk}\frac{f_{\bk}}{2}\left[\frac{\bar\e_{\bk}}{E_{\bk}}\tanh\frac{E_{\bk}}{2T}-1\right].
\eea
Without loss of generality we assume that $\bar{\Phi}_0>0$. The free energy of the coexistence state relative to the normal state is given by:
\bea\label{eq:FENem2}
\Delta \mathcal{F} &=&\mathcal{F}_{\rm nem + SC}-\mathcal{F}_{\rm normal}\nonumber\\
&=&\sum_{\bk}\left\{-T\ln\left[\frac{\cosh^2(E_{\bk}/2T)}{\cosh^2(\e_{\bk}/2T)}\right] +\bar\Phi_0f_{\bk}\right\} \nonumber\\
&&~~~~~~+ \sum_{\bk}\left\{\frac{\Delta_0^2\mathcal{Y}^2_{\bk}+\bar\Phi_0f_{\bk}\bar\e_{\bk}}{2E_{\bk}}\tanh\frac{E_{\bk}}{2T}-\frac{\bar\Phi_0f_{\bk}}{2}\right\}\nonumber\\
&=&\sum_{\bk}\left\{-T\ln\left[\frac{\cosh^2(E_{\bk}/2T)}{\cosh^2(\e_{\bk}/2T)}\right] +\bar\Phi_0f_{\bk}\right\}\nonumber\\
&&~~~~~~+\frac{\bar\Phi_0^2}{V^{\rm nem}}-\frac{\Delta_0^2}{V^{sc}}.
\eea

The free energy of the coexistence state relative to the would-be nematic state is given by
\bea\label{eq:FENem3}
\Delta \mathcal{F} &=&\mathcal{F}_{\rm nem + SC}-\mathcal{F}_{\rm nem}\nonumber\\
&=&\sum_{\bk}\left\{-T\ln\left[\frac{\cosh^2(E_{\bk}/2T)}{\cosh^2(\tilde\e_{\bk}/2T)}\right]+\left(\bar\Phi_0-\Phi_0\right)f_{\bk}\right\}\nonumber\\ 
&&~~~~~~  +\frac{\bar\Phi_0^2-\Phi_0^2}{V^{\rm nem}}-\frac{\Delta_0^2}{V^{sc}}.
\eea
Recall that $\bar\e_{\bk}=\e_{\bk}+\bar\Phi_0f_{\bk}$ is in the presence of $\Delta_0$ and $\tilde\e_{\bk}=\e_{\bk}+\Phi_0f_{\bk}$ is in the absence of $\Delta_0$ at the same $T$. For the coexistence phase to exist, we need Eq. \ref{eq:FENem3} to be negative.

\begin{figure}[htp]
$\begin{array}{c}
\includegraphics[width=0.95\columnwidth]{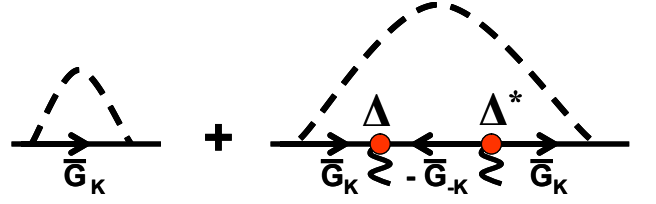}
\end{array}$
\caption{\label{fig:SE} Diagrams for {$\bar\Phi_0$}, to be expanded to order $\Delta_0^2$. {The dashed line is the nematic interaction $V^{\rm nem}$.} The $\mathcal{O}(\Delta^2_0)$ contributions come from two sources: one is directly induced by the self consistency equation (the leading order in the second diagram), and the other from the change in the nematic order parameter due to the onset of $\Delta_0$ (from $\bar G_K$ in the first diagram). There is no $\mathcal{O}(\Delta_0)$ contribution.}
\end{figure}
\subsection{Near $T_c$} Before discussing the numerical solutions to the free energy evolution with temperature and the order parameters, it is instructive to analytically consider the behavior of the self-consistent equations. This can be easily done close to $T_c$ where $\Delta_0\rightarrow 0$. Here we can set $\bar\Phi_0 \simeq \Phi_0 + \delta\Phi_0$, where $\delta\Phi_0$ is solely induced by $\Delta_0$ and is $\mathcal{O}(\Delta^2_0)$. This can be seen by expanding Eq. (\ref{eq:MF5}) in $\Delta_0$, with the leading power being $\Delta_0^2$. We further distinguish between two contributions to $\delta\Phi_0$, and label them as $\delta\Phi_0^{f}$ and $\delta\Phi_0^{\Delta}$. The former is the effect of feedback of $\Delta_0$ on to $\Phi_0$ and the latter is the ``direct" contribution of superconductivity to the self consistency equation {for $\Phi$}. This latter contribution is captured as the leading order in $\Delta_0^2$ in the second diagram of Fig. \ref{fig:SE}. Thus,
\bea\label{eq:X1}
&&\Phi_0 + \delta\Phi_0 = \Phi_0 + \delta\Phi^f_0 + \delta\Phi^{\Delta}_0\nonumber\\
&&\simeq -V^{\rm nem}\int_Kf_{\bk}\left[
-\frac{i\omega_{n}+\tilde\e_{\bk}+(\delta\Phi^f_0 + \delta\Phi^{\Delta}_0)f_{\bk}}{\omega_{n}^{2}+[\tilde\e_{\bk}+(\delta\Phi^f_0 + \delta\Phi^{\Delta}_0)f_{\bk}]^2+\Delta_0^2\mathcal{Y}^{2}_{\bk}}\right]\nonumber\\
&&\simeq -V^{\rm nem}\int_Kf_{\bk}\left[\bar G_K-G_KG_{-K}G_K\Delta_0^2\mathcal{Y}^{2}_{\bk}\right],
\eea
where $G_{\pm K}=1/(\pm i\omega_n-\tilde\e_{\pm\bk})$, $\bar G_K=1/(i\omega_n-\tilde\e_{\bk}-\delta\Phi_0f_{\bk})$, $\int_K\equiv T\sum_{n}\sum_{\bk}$, and $\bar G_K= G_K+\mathcal{O}(\Delta^2_0)$. The first term in the third line of Eq.  \ref{eq:X1} is, up to $\mathcal{O}(\Delta^2_0)$,
\bea&&-V^{\rm nem}\sum_{K}f_{\bk}\frac{1}{(i\omega_{n}-\tilde\e_{\bk})^2}[i\omega_{n}-\tilde\e_{\bk}+(\delta\Phi^f_0 + \delta\Phi^{\Delta}_0)f_{\bk}]\nonumber\\
&=&\Phi_0+\frac{(\delta\Phi^f_0 + \delta\Phi^{\Delta}_0)V^{\rm nem}}{2}\sum_{\vec{k}}f_{\bk}^2\frac{{\rm sech}^{2}[\tilde\e_{\bk}/2T]}{2T},\label{eq:X2}\eea 
whereas the second term, which is explicitly induced by the superconducting order, can be written as
\bea\delta\Phi_0^{\Delta}&=&V^{\rm nem}\sum_{\bk}\Delta_0^2\mathcal{Y}^2_{\bk}f_{\bk}\mathcal{W}_{\bk},\label{eq:X3}
\eea
where 
\bea\label{eq:X4}
\mathcal{W}_{\bk}&=&T\sum_nG_KG_{-K}G_K\nonumber\\
&=&\frac{1}{16T^2}\left[\frac1x\left\{{\rm sech}^2x-\frac{\tanh x}{x}\right\}\right],
\eea
with $x\equiv\tilde\e_{\bk}/2T$. $\delta\Phi_0^{\Delta}$ is completely determined by the system  parameters in the absence of $\Delta_0$ (i.e $\Delta_0$ is set to zero in $\mathcal{W}_{\bk}$ as shown in Eq. \ref{eq:X3}). Using Eqs. (\ref{eq:X1}), (\ref{eq:X2}) and (\ref{eq:X3}) one can eliminate $\delta\Phi_0^f$ and directly arrive at the expression for $\delta\Phi_0$ as:\bea\label{eq:X5}
\delta\Phi_0\left[1-\frac{V^{\rm nem}}{2}\sum_{\vec{k}}f_{\bk}^2\frac{{\rm sech}^{2}[\tilde\e_{\bk}/2T]}{2T}\right]&=&\delta\Phi_0^{\Delta}.\nonumber\\
\eea

\begin{figure}[htp]
$\begin{array}{c}
\includegraphics[width=0.95\columnwidth]{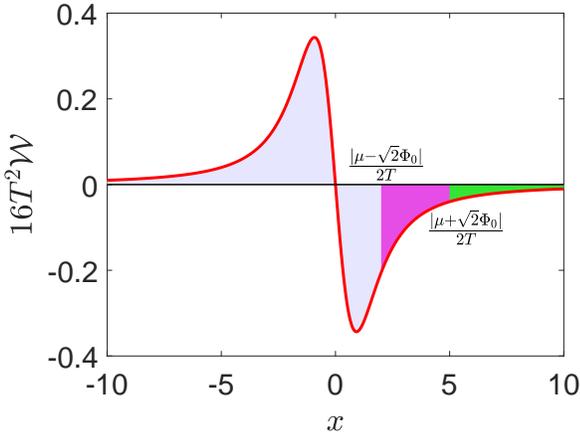}
\end{array}$
\caption{\label{fig:WW} The function $16T^2\mathcal{W}$ as a function of $x=\tilde \e_{\bk}/2T$. Because of the oddness of $\mathcal{W}$, the integration in Eq. (\ref{eq:ex}) is over the interval $[ |\mu-\Phi_0f_{\bk}|/2T$,$+\infty )$. As the angle is varied the start value of the interval itself ranges from $|\mu-\sqrt{2}\Phi_0|/2T$ to $|\mu+\sqrt{2}\Phi_0|/2T$. The largest interval is shown in magenta and the smallest one is overlaid in green.}
\end{figure}

If we define the ratio $\delta\Phi_0/\Phi_0\equiv p$, it is clear that near the onset of superconductivity, the sign of $p$ determines whether we have competition($p<0$) or cooperation($p>0$) of the two orders. We prove in the Appendix:\ref{app:c} the mathematical statement that the term in $[.]$ in Eq. (\ref{eq:X5}) is positive definite. Thus the term $\delta\Phi_0^{\Delta}$ (the ``explicit" contribution) decides whether we have competition or cooperation in the one band model.

\subsection{Competition vs cooperation in the one band model}
From Eq. \ref{eq:X3} we observe that since $V^{\rm nem}>0$, the sign of $\delta\Phi_0^{\Delta}$ is dictated by the relative anisotropy of $\mathcal{Y}_{\bk}$ and $f_{\bk}$, i.e. the interplay between the form factors of the SC gap anisotrpy and the FS distortion. This is controlled by the parameter $r$. We remind the reader that $r$ in $\mathcal{Y}_{\bk}$ should grow as $\Phi_0$ increases. However, the strength of $r$ must also be controlled by electronic structure details undetermined in this theory. We shall 
{treat $r$ as a phenomenological parameter}
and explore the phase space of allowed superconducting solutions. Changing $r$ would amount to changing the details of the electronic structure.

Returning to Eq. \ref{eq:X3},  we get
\bea\label{eq:ex}
\delta\Phi_0^{\Delta}&=&\lambda_{\rm nem}\Delta_0^2\int\frac{d\theta}{2\pi}\int_{-\mu}^{\infty} d\e_{\bk}\mathcal{Y}^2_{\bk}f_{\bk}\mathcal{W}_{\bk}\nonumber\\
&=&2T\lambda_{\rm nem}\Delta_0^2\int\frac{d\theta}{2\pi}\mathcal{Y}^2_{\theta}f_{\theta}\int^{\infty}_{(-\mu+\Phi_0f_{\theta})/2T} \mathcal{W}(x)dx.\nonumber\\
\eea

\begin{figure}[t]
$\begin{array}{c}
\includegraphics[width=0.95\columnwidth]{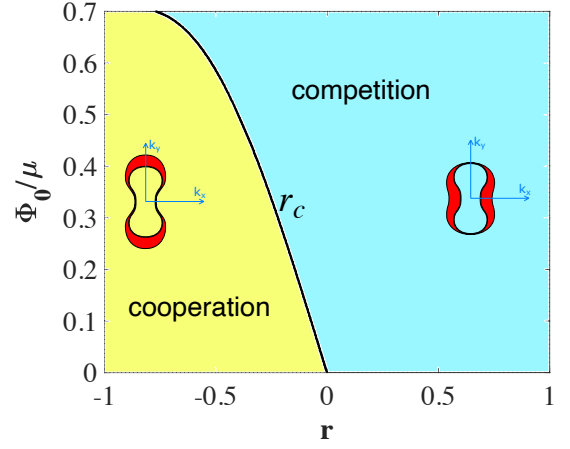}
\end{array}$
\caption{\label{fig:POM_r_phase_diagram}  The phase diagram calculated from Eq. (\ref{eq:ex2}) for the cooperative effect relating  $\Phi_0,~\mu$ and $r$. The larger  $\Phi_0$, the more negative is $r_c$, i.e. the larger is the anisotropy needed to turn the competition into cooperation.}
\end{figure}

The $1/x^2$ behavior of $\mathcal{W}(x)$ ensures that the contribution comes from around the Fermi surface, allowing us to factor the integration into radial and angular parts. While this is presented for an electron band, it can be easily extended to a hole band. Since $\mathcal{W}(x)$ is odd, the lower limit can be changed to $|\mu-\Phi_0f_{\theta}|/2T$. As shown in Fig. \ref{fig:WW}, the only surviving contribution is the tail from $|\mu-\Phi_0f_{\theta}|/2T$ to $+\infty$. In this region, assuming $(\mu-\Phi_0f_{\theta})\gg2T$, $W(x)\approx-1/16T^2x^2$ and
\bea\label{eq:ex2}
\delta\Phi_0^{\Delta}&=&-\frac{\lambda_{\rm nem}\Delta_0^2}{4}\int\frac{d\theta}{2\pi}\mathcal{Y}^2_{\theta}f_{\theta}\frac{1}{\mu-\Phi_0f_{\theta}}\nonumber\\
&=&-\frac{\sqrt{2}\lambda_{\rm nem}\Delta_0^2}{2\pi}\int_0^{\pi/2}d\theta\frac{\sqrt{2}\Phi_0(1+r^2\cos^2\theta)+2r\mu}{\mu^2-2\Phi_0^2\cos^2{\theta}}\nonumber\\
&&~~~~~~~~~~~~~~~~\times\frac{\cos^2\theta}{1+r^2/2}.
\eea

\begin{figure*}[htp]
$\begin{array}{cc}
\includegraphics[width=0.45\linewidth]{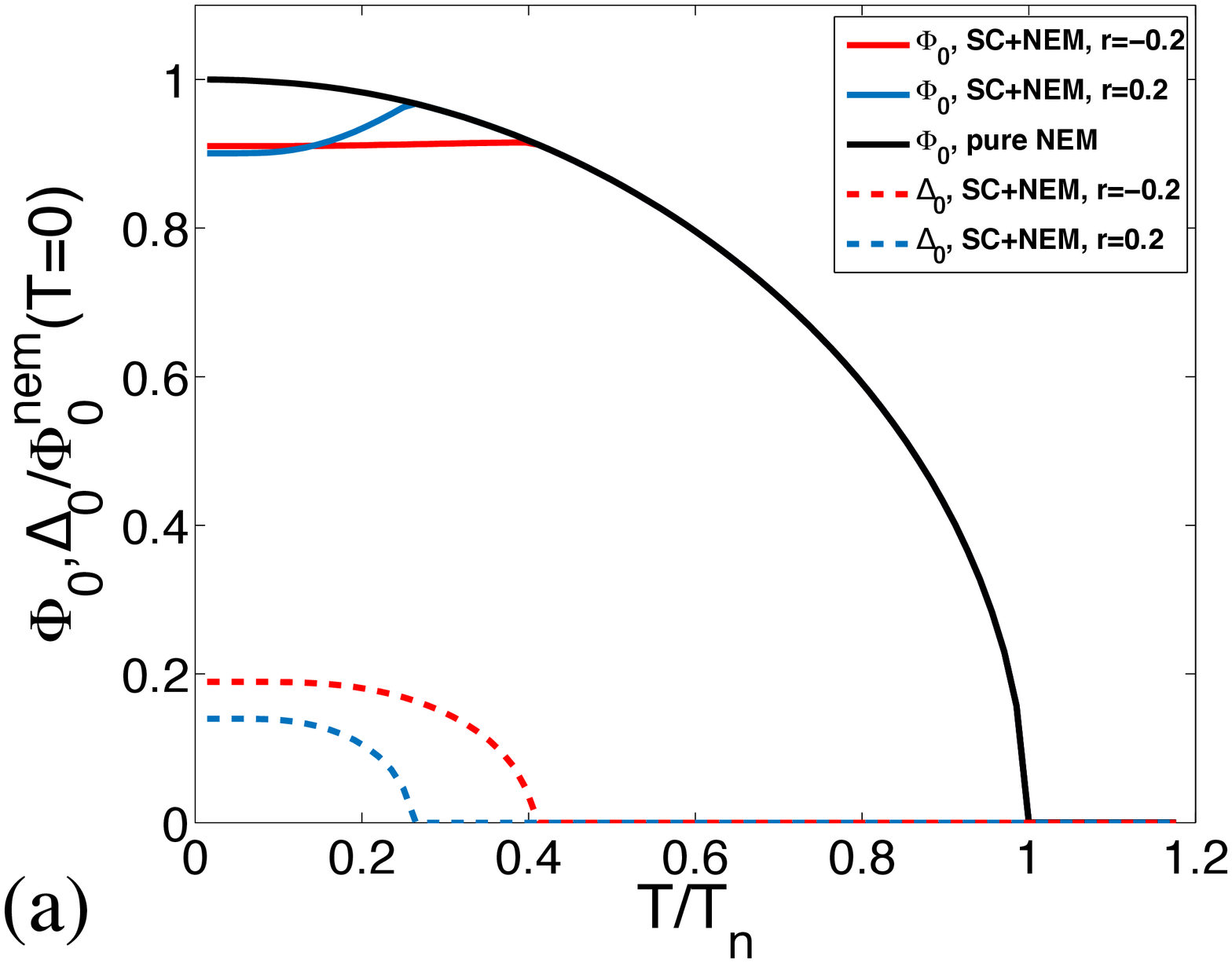}&
\includegraphics[width=0.45\linewidth]{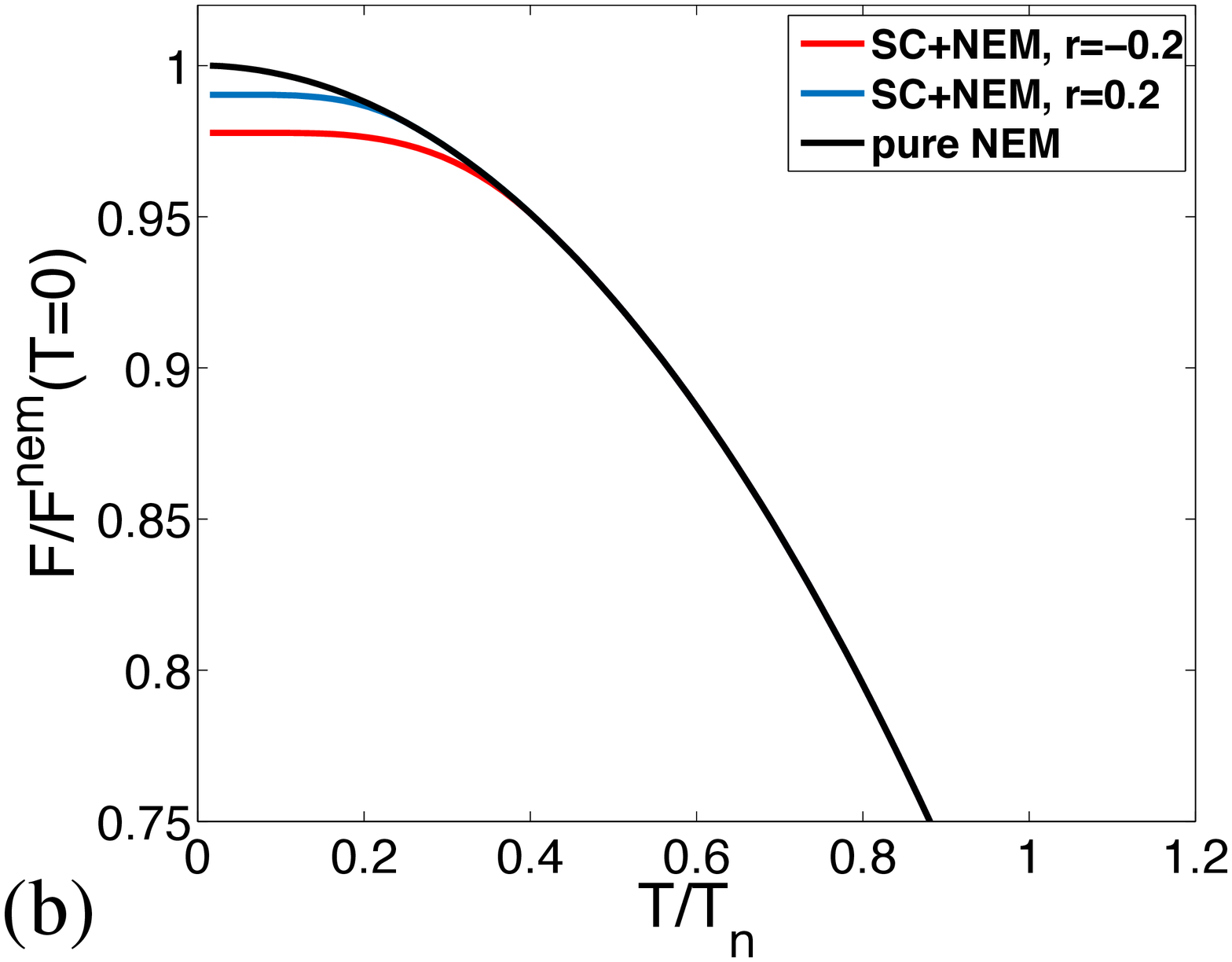}
\end{array}$
\caption{{(a)} The temperature evolution of $\Phi_0$ and $\Delta_0$ and {(b)} corresponding free energy ($\mathcal{F}$) for $r=\pm0.2$, $\lambda_{sc}\equiv \nu_0V_{sc}=-0.4$, $\lambda_{\rm nem}=1.05$, $\omega_c=2.5\mu^{nem}(T=0)$ where $\omega_c$ is the BCS cut-off energy. Here $\mu^{nem}(T=0)$ is the chemical potential of the pure nematic solution at zero temperature. 
Note that $r<0$ solution has the lower free energy.\label{fig:evo1}}
\end{figure*}

\begin{figure*}[htp]
$\begin{array}{ccc}
\includegraphics[width=0.32\linewidth]{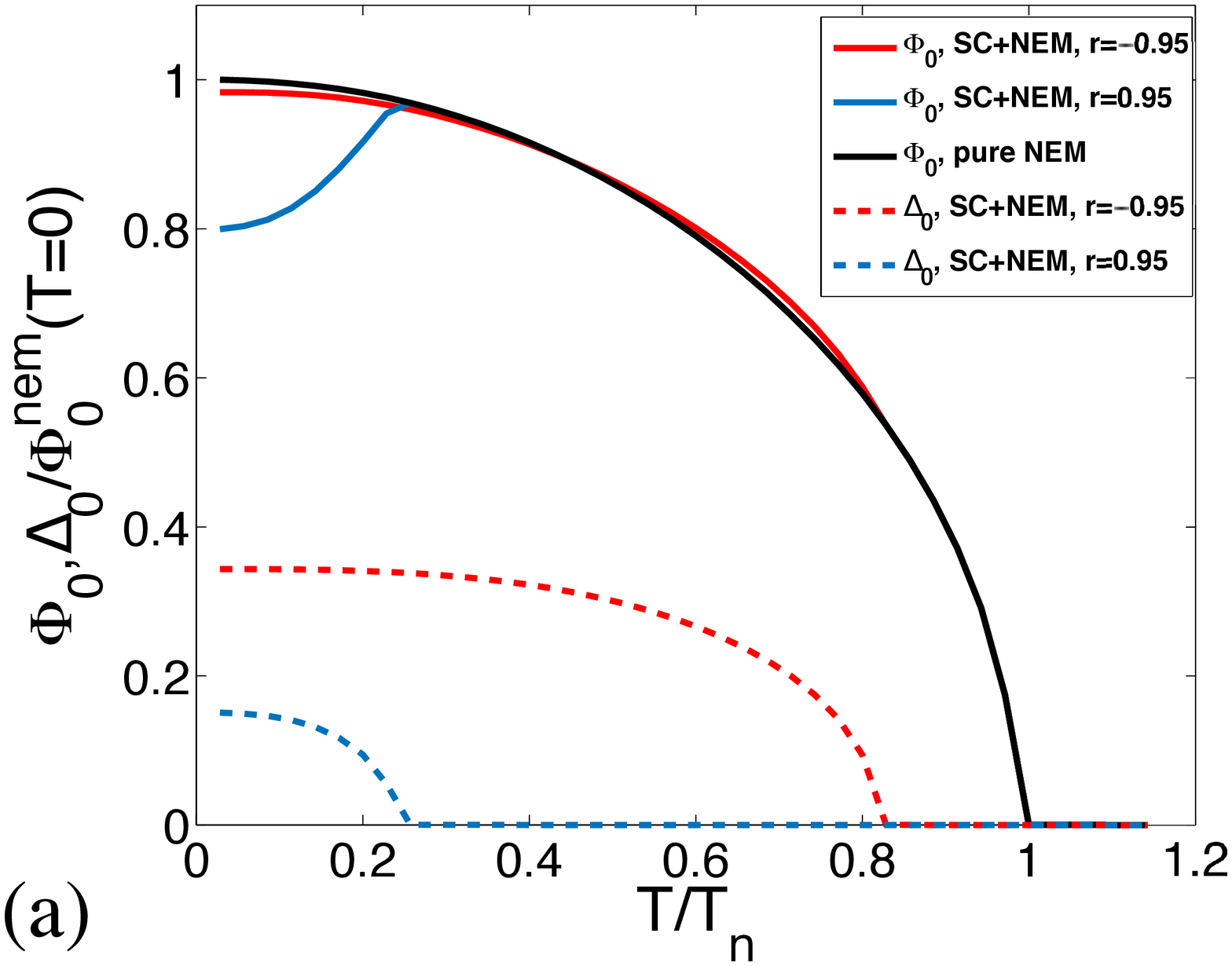}&
\includegraphics[width=0.32\linewidth]{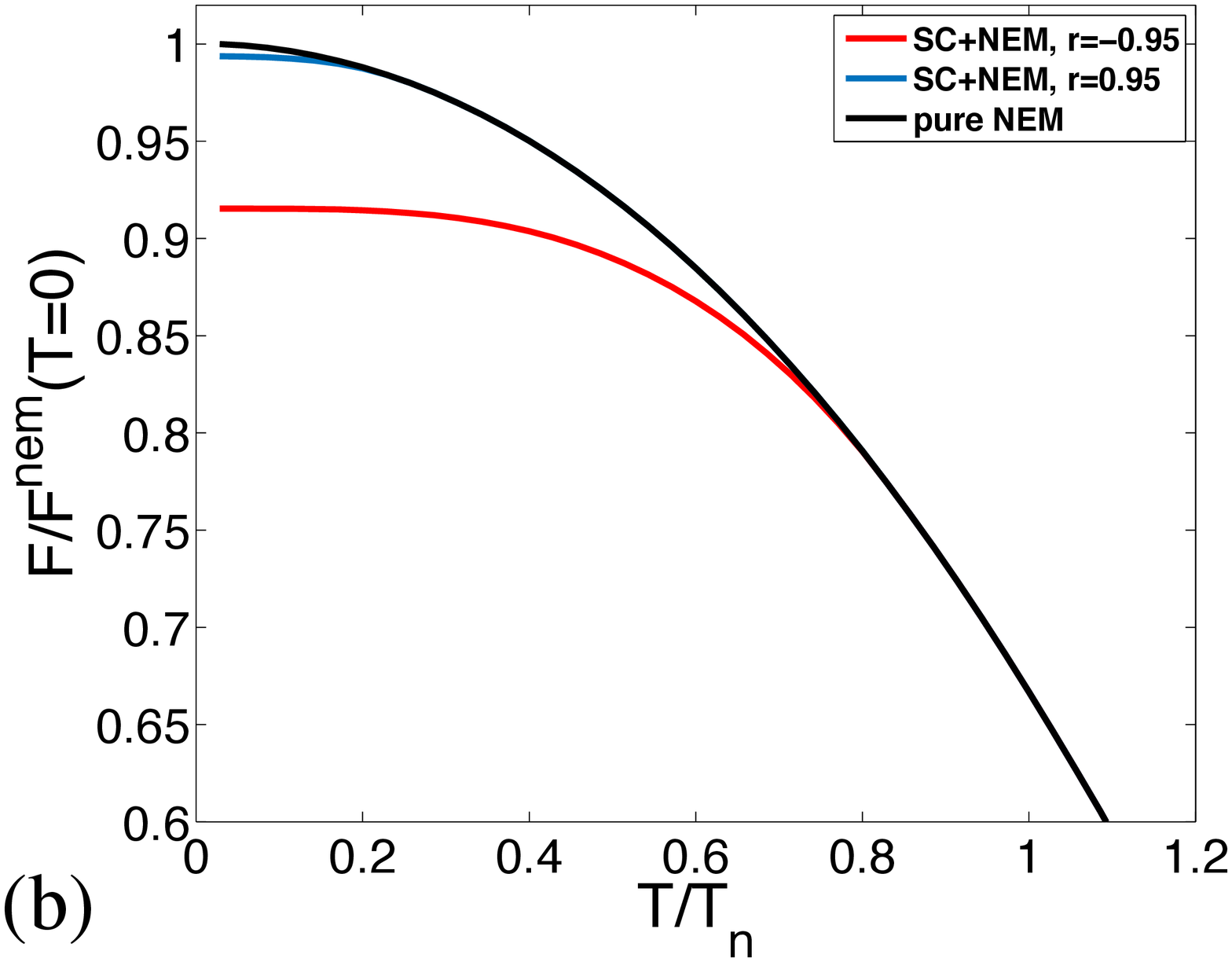}&
\includegraphics[width=0.32\linewidth]{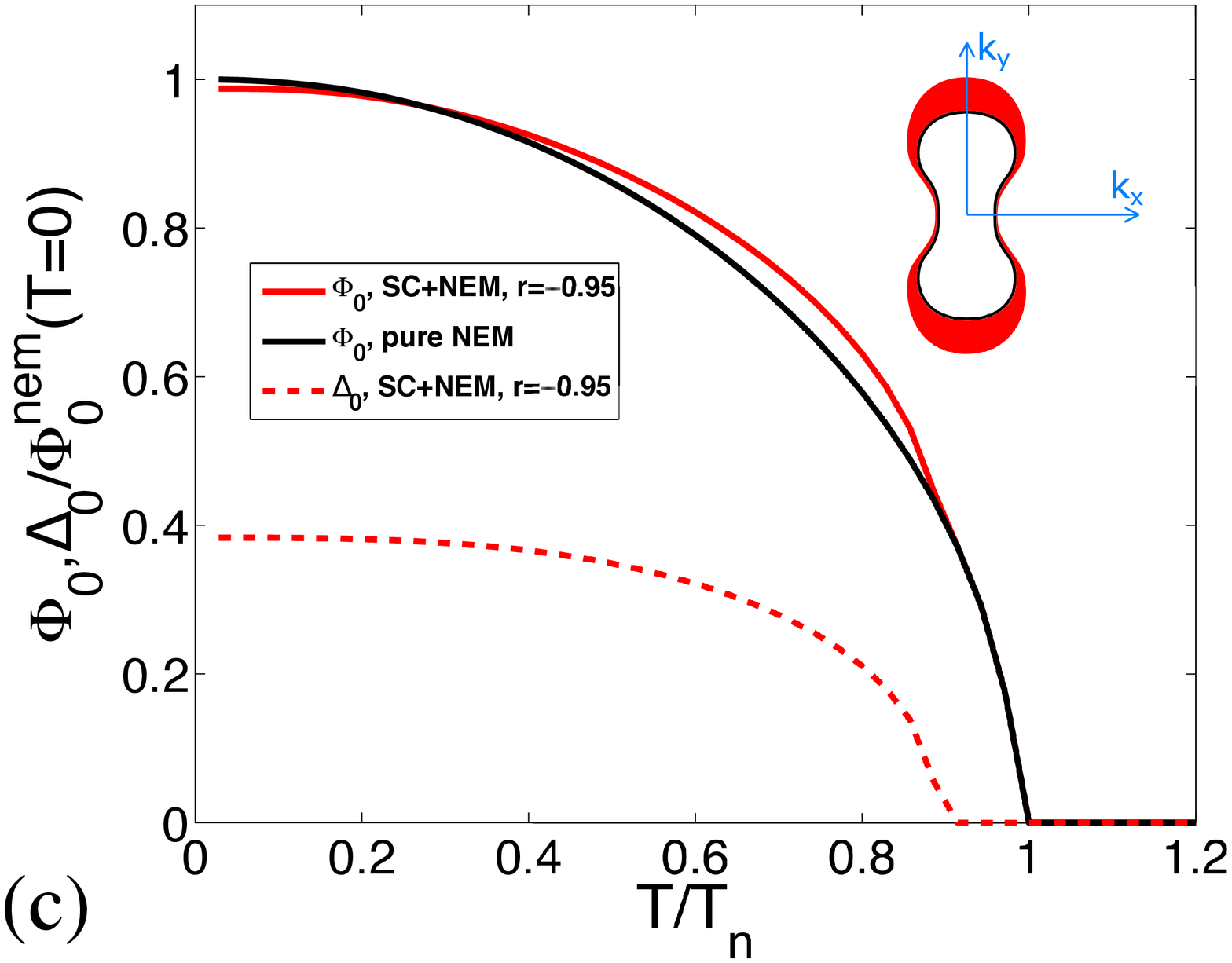}
\end{array}$
\caption{{(a)} Temperature evolution of $\Delta_0$ and $\Phi_0$ {for parameters $r=\pm0.95$, $\lambda_{sc}=-0.5$, $\lambda_{\rm nem}=1.05$, $\omega_c=2.35\mu^{\rm nem}(T=0)$.} Note the competition and cooperation for the different values of $r$. {(b)} {The free energy of the solutions in {(a)} indicating stability of the cooperating solution}. {(c)} Enhanced cooperation for parameters {$r=-0.95$, $\lambda_{sc}=-0.5$, $\lambda_{\rm nem}=1.05$, $\omega_c=2.94\mu^{\rm nem}(T=0)$. Again, this solution is more stable than that of $r=+0.95$ which shows competition. }\label{fig:evo2}}
\end{figure*}

It is clear from Eq. (\ref{eq:ex2}) that if $r\rightarrow0$, then $\delta\Phi_0^{\Delta}<0$ and hence $p<0$: this is the usual competition that is expected. If $r\neq 0$, consider first a case with a large Fermi surface ($\mu\gg\Phi_0$), then $\delta\Phi_0^{\Delta}/\Phi_0\propto -r$. This already indicates that we need $r<0$ for a cooperative effect, meaning that the Fermi-surface elongation and the gap maxima must be aligned to see  cooperation (see Fig. \ref{fig:POM_r_phase_diagram}). However, when $\mu\sim\Phi_0$ there is a threshold for $r$ beyond which cooperation  is possible. In this case, this threshold value $r_c$ is negative. It must be noted that in the former case the effect is extremely small  ($\delta\Phi_0/\Phi_0=\mathcal{O}(\Delta_0^2/\Phi_0\mu)$) due to the largeness of the Fermi-surface, thus the best case scenario to observe the cooperation effect seems to be when the Fermi surface is not too large. This indeed forms a good basis to apply such a model to FeSe.

This is the most important result for the one-band model: the correlation of the superconducting gap anisotropy with the FS elongation (due to nematic order) seems to affect the competition vs cooperation outcome. More specifically, it demonstrates that if $r>0$ (``anti-aligned" gap and FS elongation), we always have competition. If $r<0$, there is a critical negative $r_c$ beyond which $p$ reverses sign changing the more common competition to cooperation. {This result applies beyond the GL regime.}

Let us now move to the temperature dependent numerical solutions to the self consistent equations (Eqs. \ref{eq:MF4a}, \ref{eq:MF5}) which are solved together with the self-consistent determination of $\mu$  from a fixed total number of particles. We also demonstrate the stability of these solutions by analysing the free energy. Figure \ref{fig:evo1} demonstrates the usual competition for values of $r$ that are above the threshold anisotropy. Note that  negative $r$ corresponds to the more stable solution.  In Fig. \ref{fig:evo2} (a) and (b), we demonstrate the cooperative effect for $r=-0.95$ (below the threshold anisotropy). Note that changing the sign of $r$ removes the cooperative effect; and that the cooperative solution is the more stable one. In Fig. \ref{fig:evo2}{(c),} we demonstrate enhanced cooperative effect for slightly different parameters that enhance the superconducting transition temperature.

\begin{figure}[htp]
$\begin{array}{cc}
\includegraphics[width=0.95\linewidth]{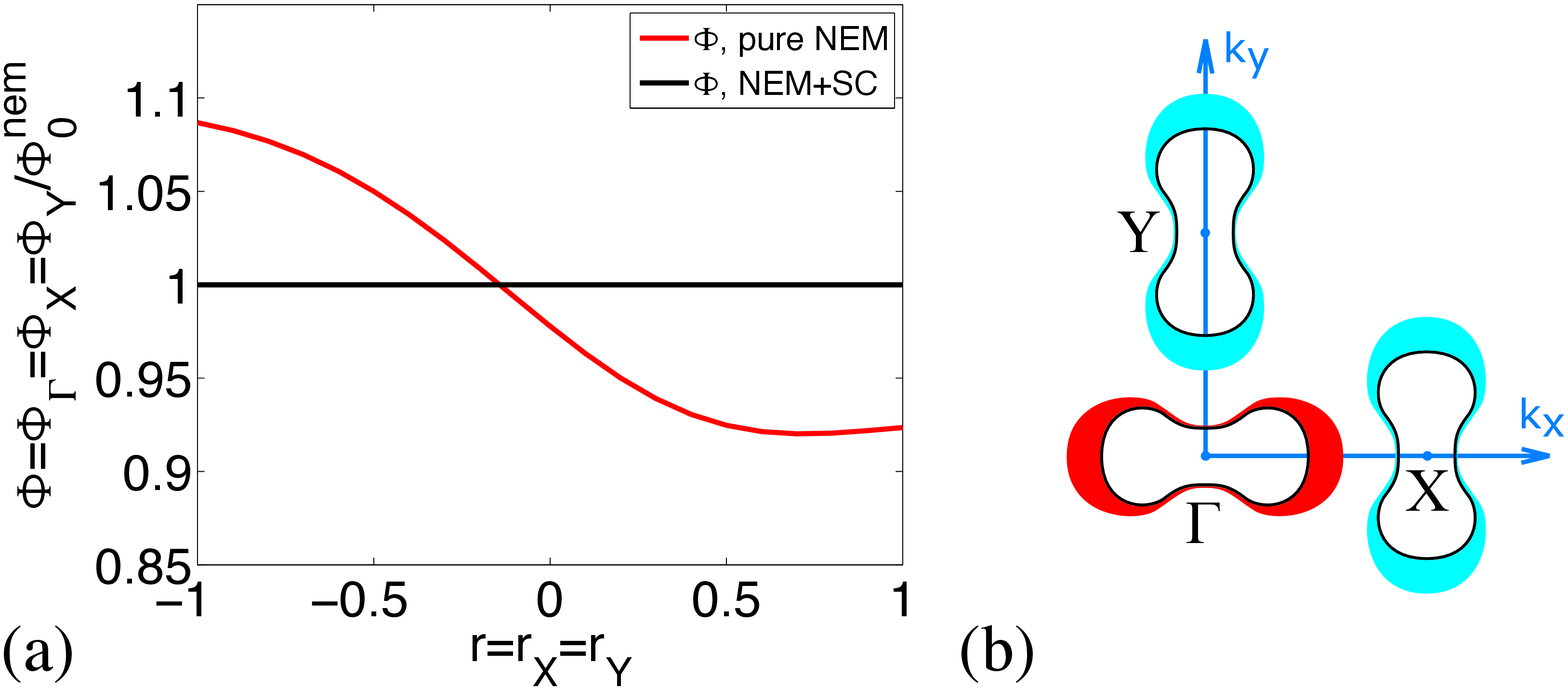}
\end{array}$
\caption{\label{fig:3bandPhiatT0} {(a)} Solutions of $\Phi's$ at $T=0$ as functions of $r=r_X=r_Y$ for parameters $\varepsilon=0$, $\alpha=0$, $\beta=1$, all $\lambda^{nem}_{ab}=0.425$, $\lambda_{{\Gamma}X}^{sc}=\lambda_{{\Gamma}Y}^{sc}=0.4$, $\lambda_{\Gamma\Gamma}^{sc}=\lambda_{XX}^{sc}=\lambda_{YY}^{sc}=\lambda_{XY}^{sc}=0$,  with $\lambda^{{\rm nem},sc}_{ab}=V^{{\rm nem},sc}_{ab}\nu_{0}$, $r_{\Gamma}=1$ and $\mu_h=\mu_e=0.2\omega_c$. The red curve is the stable solution, which coexists with a superconducting order, where the black line is the solution without  superconductivity. At large negative $r$, enhanced nematic order due to superconductivity is found. {(b)} A sketch showing Fermi surface elongation(black solid contour) and magnitude of superconducting gap around each pocket(width of colored region) of the cooperative solution in {(a)} at large negative $r$. Different colors on electron and hole pockets mean a sign reversal of the gap. This cooperative solution has, on each pocket, the gap maximum and the elongation of Fermi surface contour in the same direction.}
\end{figure}

\begin{figure*}[htp]
$\begin{array}{cc}
\includegraphics[width=0.45\linewidth]{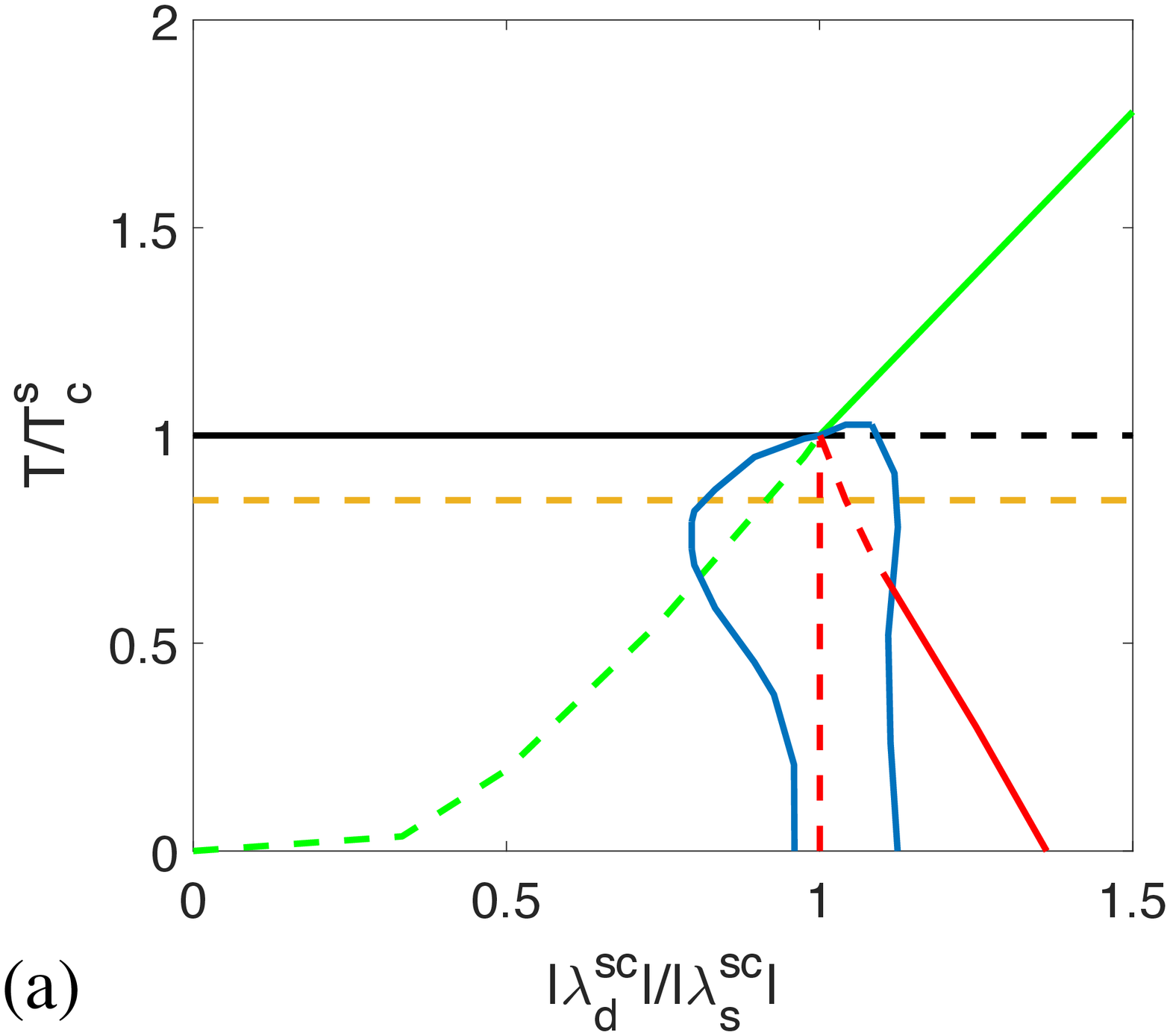}
\includegraphics[width=0.45\linewidth]{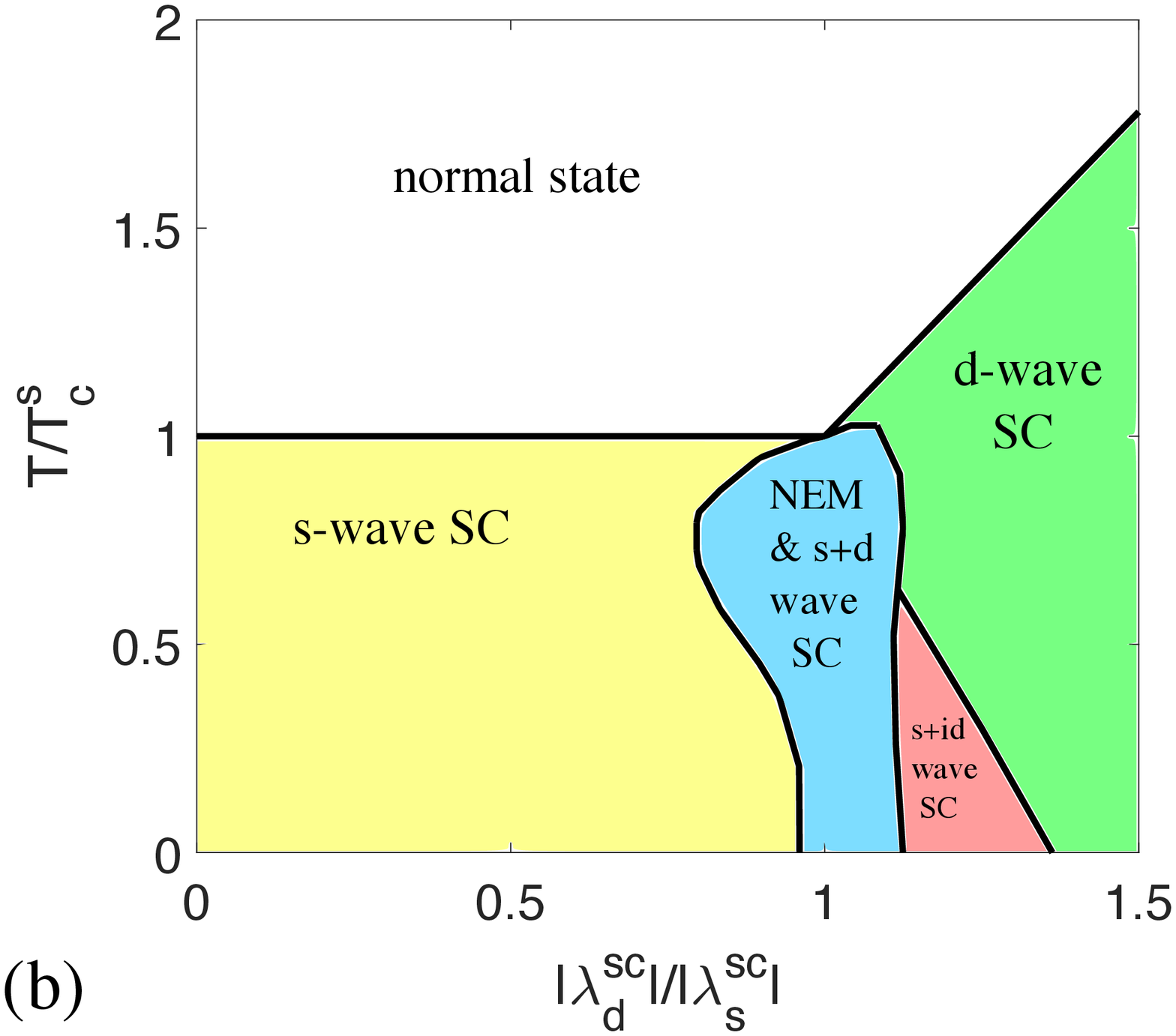}&
\end{array}$
\caption{\label{fig:T_c>T_n_phasediagram} {(a)} Transition temperatures of different solutions of our $T_c>T_n$ model with parameters $\lambda^{sc}_{s}=-0.6$, $\lambda^{\rm nem}=1.04$, $\omega_c=2.92\mu^{\rm nem}(T=0)$. Black and green lines are $T_c$'s of the pure $s$ and the pure $d$-wave superconducting solutions respectively, with their solid portions indicating the leading instability of the system. The yellow dashed line is the onset of the pure nematic solution. The nematic solution that develops out of a preexisting $s$ or $d$ wave superconducting order and coexists with an $s+d$ wave superconductivity is enclosed by the blue curve.  The solid red line separates the $s+id$ state from the $d$-wave state, and the dashed red line represents the boundary of the $s+id$ state in the absence of nematicity.  {(b)} The phase diagram consisting of only the actual transitions from {(a)} confirmed by free energy calculation.}
\end{figure*}

Although the discussion above involved an electron band, the results for a hole band are the same. The equations can be obtained by $m\rightarrow-m,~\mu\rightarrow-\mu,~\Phi_0\rightarrow-\Phi_0$ and $r\rightarrow-r$. In particular, the conclusion that the  Fermi-surface  elongation  and  the  gap  maxima must be in the same direction for  cooperation to take place is also valid in the case of a hole band.

\subsection{Effect of multiplicity of bands}
Here we quickly demonstrate that cooperation can also occur when the system has multiple bands. A minimal model for the Hamiltonian for a system with a hole pocket at $\Gamma$ and an electron pocket each at the $X/Y$ points (without any orbital characterization) can be written as:
\bea\label{eq:3BandH}
H&=&\sum_{\bk sa}\epsilon_{\bk}^a c_{\bk sa}^{\dagger}c_{\bk s a}+H_{\rm int}^{\rm SC} + H_{\rm int}^{\rm nem},\nonumber\\
H_{\rm int}^{\rm nem}&=&-\frac14\sum_{ab\bk\bk'ss'}V_{a\bk b\bk'}^{\rm nem}
c_{\bk s a}^{\dagger}c_{\bk s a}c_{\bk's' b}^{\dagger}c_{\bk's'b},\nonumber\\
H_{\rm int}^{\rm SC}&=&\frac14\sum_{ab\bk\bk'ss'tt'}V_{a\bk b\bk'}^{sc}c_{\bk s a}^{\dagger}c_{-\bk s'a}^{\dagger}c_{-\bk'tb}c_{\bk't'b}\sigma^y_{ss'}\sigma^y_{tt'},\nonumber\\
\eea
where $a,b\in\{\Gamma,X,Y\}$, and normal band dispersions $\epsilon_{\bk}^a$ are $$\epsilon_{\vec{k}}^{\Gamma}=\mu_{h}-\frac{k^{2}}{2m},$$ $$\epsilon_{\vec{k}}^{X}=\frac{k_{x}^{2}}{2m(1+\varepsilon)}+\frac{k_{y}^{2}}{2m(1-\varepsilon)}-\mu_{e},$$
$$\epsilon_{\vec{k}}^{Y}=\frac{k_{x}^{2}}{2m(1-\varepsilon)}+\frac{k_{y}^{2}}{2m(1+\varepsilon)}-\mu_{e},$$ where $\bk$ of each band is measured from the corresponding center of the pocket, and $\varepsilon<1$ is a parameter controlling the ellipticity of the electron pockets. The interactions take the factorized form 
\begin{equation}
V^{\rm nem}_{a\bk b\bk'}=V^{\rm nem}_{ab}g_a(\bk)g_b(\bk'),
\end{equation}
and 
\begin{equation}
V^{sc}_{a\bk b\bk'}=V^{sc}_{ab}{{\cal Y}_a(\vec k){\cal Y}_b(\vec k')},
\end{equation}
where 
\begin{eqnarray}
g_{\Gamma}(\bk)&=&\sqrt{2}\cos2\theta_{\bk},\\ g_{X}(\bk)&=&\left(\alpha+\beta\cos2\theta_{\bk}\right)/\sqrt{\alpha^2+\beta^2/2},\\ g_{Y}(\bk)&=&\left(-\alpha+\beta\cos2\theta_{\bk}\right)/\sqrt{\alpha^2+\beta^2/2},\\  {\cal Y}_{a}(\vec k)&=&(1+r_{a}\cos2\theta_{\bk})/\sqrt{1+r_{a}^2/2}. 
\end{eqnarray}
Note that all $\theta_{\vec k}$'s are measured with respect to the $x$ axis. Since $\Delta_X$ and $\Delta_Y$ are in general different in the nematic phase, the form factor ${\cal Y}_{a}$ enables general form of $s+d$ wave gaps over the whole Brillouin zone, with angular harmonics up to $\cos2\theta_{\bk}$ on each pocket.

Proceeding with the mean-field approximation as before, we get
\bea
H_{MF}=\underset{\vec{k},s,a}{\sum}[\epsilon_{\vec{k}}^{a}+\Phi_{a}g_{a}(\vec{k})]c_{\vec{k}sa}^{\dagger}c_{\vec{k}sa}\nonumber\\
-\underset{\vec{k},a}{\sum}\left(\Delta_{a}{\cal Y}_{a}(\vec{k})c_{\vec{k}\uparrow a}^{\dagger}c_{-\vec{k}\downarrow a}^{\dagger}+h.c.\right),
\eea
with nematic and superconducting order parameters 
\bea
\Phi_{a}&=&\underset{b,\vec{k'},s'}{\sum}-\frac{1}{2}V_{ab}^{\rm nem}g_{b}(\vec{k'})<c_{\vec{k'}s'b}^{\dagger}c_{\vec{k'}s'b}>,\\
\Delta_{a}&=&-\underset{\vec{k'},b}{\sum}V_{ab}^{sc}{\cal Y}_{b}(\vec{k}')<c_{-\vec{k}'\downarrow b}c_{\vec{k'}\uparrow b}>.
\eea
We self-consistently solve for nematicity and superconductivity just as in the one-band case. The behavior is not universal, as there are many parameters for the electronic dispersion and interactions. Nevertheless we are able to demonstrate a possible case of cooperation in such systems as shown in Fig. \ref{fig:3bandPhiatT0}.  For simplicity, the parameters here have been chosen (see caption for parameters) such that the nematic order parameters $\Phi_{a}$ on all pockets are equal to $\Phi$. Fig. \ref{fig:3bandPhiatT0}{(a)} shows solutions of $\Phi_{\Gamma}=\Phi_{X}=\Phi_{Y}\equiv\Phi$ at $T=0$ as functions of $r_X=r_Y {\equiv} r$ that controls the gap anisotropy on the electron pockets. The red curve is the solution with the minimum free energy, which also coexists with  superconducting order, while the black horizontal line is the pure nematic solution $\Phi_{a}=\Phi^{\rm nem}_0$. At large negative $r$, enhanced $\Phi$ due to the onset of a superconducting order is observed. The cooperation between nematicity and superconductivity results in a state such that on each pocket the gap maximum and the elongation of the Fermi surface are in the same direction, as sketched in Fig. \ref{fig:3bandPhiatT0}{ (b)}.

\begin{figure*}[htp]
$\begin{array}{cc}
\includegraphics[width=0.95\linewidth]{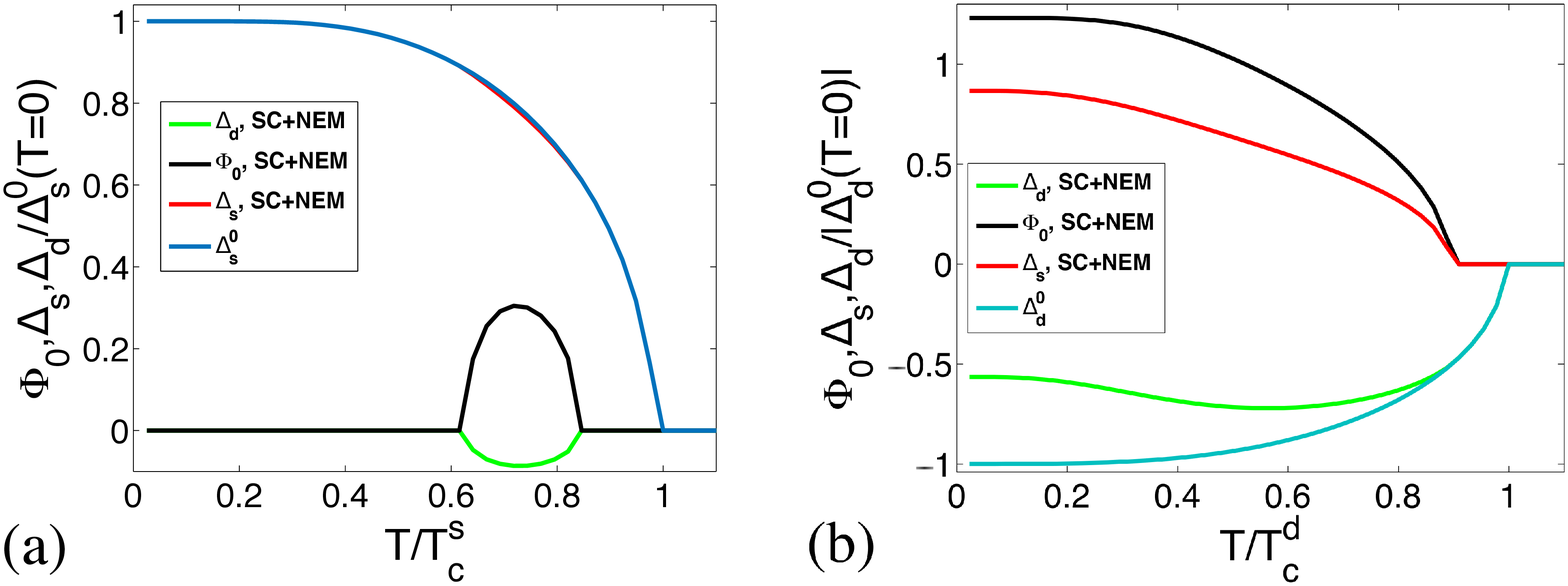}
\end{array}$
\caption{\label{fig:OPs_left_side_in_Phase diagram} Order parameters as functions of temperature for {(a)} $|\lambda_d^{sc}|=0.82|\lambda_s^{sc}|$ and {(b)} $|\lambda_d^{sc}|=1.08|\lambda_s^{sc}|$ in Fig. (\ref{fig:T_c>T_n_phasediagram}){(a)}. Here $\Delta_s^0$ and $\Delta_d^0$ are the pure $s$ and the pure $d$-wave superconducting solutions respectively.}
\end{figure*}

\section{Scenario 2: $T_n<T_c$}\label{sec:B}
The preceding discussion was based on the assumption that superconductivity condenses inside the nematic phase, which is indeed the case in many Fe-based systems where superconductivity and nematic order coexist. However, other situations exist and are interesting.   For example, when FeSe is doped with S\cite{Meingast_S-dopedFeSe}, the nematic phase transition line apparently crosses the superconducting dome, such that a transition from a tetragonal superconductor to a nematic one should be in principle observable: for a narrow range of S concentrations,  $0<T_n<T_c$. Similar crossings  take place in the phase diagrams of Co-doped NaFeAs and BaFe$_2$As$_2$.
More recently, low-$T$ ARPES data in tetragonal  LiFeAs indicated a $C_4$ symmetry breaking of the superconducting gap function below $T_c$, although the transition point itself was not determined\cite{Kushnirenko2018}.  

To study these and related cases, we propose the scenario (b) of Sec. \ref{sec:model} where nematicity coexisting with superconductivity is described by two competing attractive channels, with $s$ and $d$ symmetry.  In the tetragonal phase, spin fluctuations models of electron pairing in  Fe-based systems have shown that these two channels may closely compete\cite{Graser2009,Maiti2011}.  In the absence of nematic order, however, the well-known weak-coupling solution to the problem\cite{Musaelian1996} shows that only pure $s$, pure $d$, or $s+id$ solutions are energetically favorable; all of these will have $C_4$ symmetric quasiparticle spectra and energy gaps. We show below that it is possible  for the system to spontaneously break tetragonal symmetry at $T_n\le T_c$, however.  Special cases of these solutions were found in earlier studies\cite{Livanas2015,Fernandes_Millis_PRL2013,Watashige2015}, and shown to be either  real (``$s+d$") or complex with internal phase generally different from $\pi/2$ (``$s+e^{i\theta}d$"), depending on details of the system.  

The model Hamiltonian takes the form of Eqs. (\ref{eq:starting H}) and (\ref{eq:start}). The superconducting gap is expressed as a sum of $s$ and $d$ wave harmonics $\Delta_{\bk}=\Delta_s+\Delta_df_{\bk}$. The mean-field self-consistency equations of the order parameters read
\bea\label{eq:T_c>T_n_OPeqs}
\Phi_0&=& -V^{\rm nem}\sum_{\bk}f_{\bk}\langle c^\dagger_{\bk}c_{\bk}\rangle,\\
\Delta_s&=& -V^{s}\sum_{\bk}\langle c_{-\bk\downarrow}c_{\bk\uparrow}\rangle,\\
\Delta_d&=& -V^{d}\sum_{\bk}f_{\bk}\langle c_{-\bk\downarrow}c_{\bk\uparrow}\rangle.
\eea
Letting $\lambda^{sc}_{s,d}=\nu_0V^{s,d}$, $\lambda^{\rm nem}=\nu_0V^{\rm nem}$,  we look for solutions of the type $\Delta_s + e^{i\theta}\Delta_d$. For a fixed $\lambda^{sc}_s$, we obtain the phase diagram in the $T-\lambda^{sc}_{d}$ plane as shown in Fig. (\ref{fig:T_c>T_n_phasediagram}).

Fig. (\ref{fig:T_c>T_n_phasediagram}){(a)} shows the transition temperatures of different solutions of the model. The black and the  green lines are onset temperatures of pure $s$ and pure $d$ wave superconducting solutions respectively, with their solid parts indicating the leading instability for the corresponding $\lambda^{sc}_{d}$,  and the dashed ones indicating the subleading one. The  blue solid curve is the phase boundary of the nematicity that develops out of a preexisting $s$ or $d$-wave superconducting order and coexists with an $s+d$ wave superconductivity. The yellow dashed line represents the onset temperature of the pure nematic solution.  The red solid line marks the boundary between $d$-wave and the $s+id$ phase. The red dashed line indicates the boundary of the $s+id$ phase that would exist in a system without any nematic order. We note that we do not find a solution which is of the form $s+e^{i\theta}d +\Phi_0$, with $\theta\neq 0$ or $\pi$. Free energy calculations confirm that the actual transitions take place only at the solid lines, yielding the phase diagram as in Fig. (\ref{fig:T_c>T_n_phasediagram}){(b)}. All solid lines represent second order phase transitions except the boundary between the $s+id$ and $s+d+\Phi$ phases, where the transition  is discontinuous. 

Here it is important to notice that the nematic phase exists only  around $|\lambda_d^{sc}|/|\lambda_s^{sc}|\approx1$, i.e. where the $s$ and the $d$ wave superconducting channels are nearly degenerate. At the exact degenerate point, $T_n=T_c$, and $T_n$ is enhanced from the onset temperature of the pure nematic solution due to the coexisting $s+d$ superconductivity. Re-entrance behavior is also observed in a narrow region of $|\lambda_d^{sc}|$ to the left of the degenerate point. In this region, with decreasing temperature, the system first enters the nematic phase from a pre-developed $s$-wave superconductivity, then leaves this phase at a lower temperature as shown in Fig. (\ref{fig:OPs_left_side_in_Phase diagram}){(a)}. Fig. (\ref{fig:OPs_left_side_in_Phase diagram}){(b)} displays the onset of the nematic order and the $s$-wave gap inside a $d$-wave superconducting state to the right of the degenerate point.

\section{Our results in the context of experiments}\label{sec:exp}
While the interplay of nematicity and superconductivity has been investigated before in the GL formalism\cite{Fernandes_Millis_PRL2013,Kushnirenko2018}, this work considers a microscopic model  provides a benchmark for further investigations. 
The cooperative effect reported here in both the 1-band and 3-band cases is consistent with the thermodynamic data in Ref. \onlinecite{Meingast_S-dopedFeSe} on FeSe$_{1-x}$S$_{x}$.  However, in this cooperative case, note that both our  1-band and 3-band model results predict the gap anisotropy to align with the FS elongation.  In fact, the gap structure in FeSe$_{1-x}$S$_{x}$ reported by ARPES in Ref. \onlinecite{Xu_DLFeng2016} is anti-aligned.  Furthermore, a similar contradiction with the measured gap structure of FeSe itself\cite{Sprau2017} in a calculation with a similar model was reported in Ref. \onlinecite{Mishra_NJP2016}, where the observed rise of $T_c$ upon electron irradiation\cite{Teknowijoyo16} was found to require competition of nematic and superconducting order. These two discrepancies are almost certainly an indication that orbital physics may be relevant to  observe the cooperative effect  with anti-aligned distortion, since the momentum dependence of the interaction in the current model is
taken as given, and is thus equivalent to a band-only model where the interaction depends exclusively on the angle-dependent density of states.  The presence of mixed orbital character in states near a given Fermi surface sheet can, via trivial matrix element effects or via many-body decoherence, create a dramatically different momentum dependence than that expected from DOS effects, e.g. nesting. A study of competition vs. cooperation of nematicity and superconductivity  including these factors will be part of future investigations.


Recently it was reported that LiFeAs may be a nematic superconductor\cite{Kushnirenko2018}, breaking the tetragonal symmetry of the normal state at some temperature below $T_c$.  Within our framework, such a result is quite possible, especially if there is a competing $d$-wave channel.  It would be interesting to seek independent evidence for the existence of competing superconducting channels, e.g. the existence of Bardasis Schrieffer type modes in the Raman spectrum\cite{Bohm2014,Maiti2016}.  To our knowledge,  measurements of electronic Raman scattering below $T_c$  capable of detecting such modes have not been reported on LiFeAs. It is worth noting that the competition and cooperation effects we discussed in this article, along with the phase diagram in Fig. \ref{fig:T_c>T_n_phasediagram}, provide insights into the possible phases that result from the interplay between superconductivity and nematicity. In the phase diagram of Fig. \ref{fig:T_c>T_n_phasediagram}, we kept the electronic occupation fixed and used the ratio between the $d$-wave and $s$-wave coupling constants as the independent tuning parameter. Determining its relationship to typical experimental tuning parameters, such as chemical substitution and pressure, is a challenging task that depends crucially on microscopic considerations. While this is left for a future project, we note that in the relevant case of S-doped FeSe, the electronic occupation is unchanged, since S is isovalent to Se.

\section{Conclusion}\label{sec:Conclusion}
In this work, we have presented a  model that allows us to microscopically study whether  superconductivity and nematicity compete or cooperate. While the former is the more common and expected scenario, this work shows that for certain anisotropic pairing interactions cooperation is also possible.  In our current model, where orbital degrees of freedom are neglected,
a signature of the cooperation would be the alignment of the FS  elongation with the superconducting gap-maxima.   We note that the comparison with a recent experiment on the S-doped FeSe system, that exhibits cooperation of nematic and superconducting orders, appears to show the  opposite orientation of the gap maxima relative to the Fermi surface distortion, leading us to believe that the orbital effects neglected here play a crucial role in these systems.

We have verified our conclusions for both   1- and 3-band models.  Interestingly, although we were not able to explore the parameter space of the 3-band model thoroughly, we find that cooperation appears to be significantly more likely to occur, and stronger than in 1-band systems. We have also shown that if nematicity emerges from superconductivity, the cooperation is still seen when the superconducting state has competing $s$- and $d-$wave orders, of possible relevance to recent measurements on LiFeAs. We note that the cooperative effect is diminished at lower temperatures.

Our results open up some obvious new lines of inquiry. Having thoroughly understood the one-band results, one can use this to study the effect of multiple orbitals making up the band, and  study the effect of disorder on this phenomenon.  Full exploration of the phase space for 3 band models is also called for. More ambitious still, will be inclusion of a pairing interaction that is derived from electronic scattering processes, e.g. spin fluctuations, based on the underlying, distorted Nematic band structure as it evolves with temperature\cite{Kang2014}. Studies along these lines are ongoing.

\vskip .2cm
{\bf Acknowledgements.}  The authors are grateful for useful conversations with C. Meingast and I.I. Mazin. XC and  PJH were partially supported by the US Department of Energy  under award DE-FG02-02ER45995. SM was at UF when the project began and acknowledges support of the Natural Sciences and Engineering Research Council of Canada (NSERC) [RGPIN-2019-05486]. RMF was supported by the US Department of Energy under award DE-SC0020045.

\bibliography{main}

\begin{center}
{\bf Appendix}
\end{center}
  
\subsection{Projecting out the Pomeranchuk and singlet superconducting channels}\label{app:a}
Consider the following re-writing of the interaction term
\bea\label{eq:app1}
H_{\rm int}&=&\frac12\sum_{\bq}V(\bq)n(\bq)n(-\bq)\nonumber\\
&=&\frac12\sum_{\bk\bk'\bq}V(\bq)c^{\dagger}_{\bk\alpha}c^{\dagger}_{\bk'\beta}
c_{\bk'+\bq\gamma}c_{\bk-\bq\delta}
\delta_{\alpha\delta}\delta_{\beta\gamma}\nonumber\\
&=&-\frac14\sum_{\bk\bk'\bq}V(\bk-\bk')
c^{\dagger}_{\bk\alpha}c_{\bk+\bq\beta}
c^{\dagger}_{\bk'+\bq\gamma}c_{\bk'\delta}\nonumber\\
&&~~~~~~~~~~~~~~~~~\times\left[\delta_{\alpha\beta}\delta_{\gamma\delta}+\vec\sigma_{\alpha\beta}\cdot\vec\sigma_{\gamma\delta}\right]\nonumber\\
&=&-\frac14\sum_{\bk\bk'\bq}V^{nm}f^*_{n}(\bk)f_m(\bk')
c^{\dagger}_{\bk\alpha}c_{\bk+\bq\beta}
c^{\dagger}_{\bk'+\bq\gamma}c_{\bk'\delta}\nonumber\\
&&~~~~~~~~~~~~~~~~~\times\left[\delta_{\alpha\beta}\delta_{\gamma\delta}+\vec\sigma_{\alpha\beta}\cdot\vec\sigma_{\gamma\delta}\right].
\eea
In an inversion symmetric system in the continuum limit, $V^{nm}\rightarrow V^n\delta_{nm}$. Under our assumption, we expect the above bare interaction term to grow such that $d-$wave charge channel (the term with $\delta_{\alpha\beta}\delta_{\gamma\delta}$ and $n=2$) to be relevant over the other terms. The instability is expected at $q=0$ as the static susceptibility is peaked at $q=0$. Picking this $\bq$ we arrive at $H_{\rm int}^{\rm Nem}$. We denote the renormalized interaction in this channel with $V^{\rm nem}$.

Similarly, we can investigate the Cooper channel by re-writing the interaction term as:
\bea\label{eq:app2}
H_{\rm int}&=&\frac12\sum_{\bq}V(\bq)n(\bq)n(-\bq)\nonumber\\
&=&\frac14\sum_{\bk\bk'\bq}V(\bk-\bk')
c^{\dagger}_{\bk\alpha}c^{\dagger}_{-\bk+\bq\beta}
c_{-\bk'+\bq\gamma}c_{\bk'\delta}\nonumber\\
&&~~~~~~~~~~~~~~~~~\times\left[\delta_{\alpha\beta}\delta_{\gamma\delta}+\vec\sigma_{\alpha\beta}\cdot\vec\sigma_{\gamma\delta}\right]\nonumber\\
&=&\frac14\sum_{\bk\bk'\bq}V^{nm}f^*_n(\bk)f_m(\bk')
c^{\dagger}_{\bk\alpha}c^{\dagger}_{-\bk+\bq\beta}
c_{-\bk'+\bq\gamma}c_{\bk'\delta}\nonumber\\
&&~~~~~~~~~~~~~~~~~\times\left[\delta_{\alpha\beta}\delta_{\gamma\delta}+\vec\sigma_{\alpha\beta}\cdot\vec\sigma_{\gamma\delta}\right].
\eea
Here we assume the singlet channel $\sigma^y$ for $n=0=m$ is enhanced over the triplet and other singlet channels. Condensation happens at $q=0$ because the Cooper logarithm is the strongest at $q=0$. Setting $\bq=0$, we are led to $H^{\rm SC}_{\rm int}$. We denote the renormalized interaction in this channel with $V^{sc}$. It should be noted that there is no double counting involved since the components of $V$ that are enhanced correspond to different processes (particle-hole scattering for nematic and particle-particle scattering for superconducting). Different interaction matrix elements contribute to these processes and can thus be separately enhanced.

\subsection{Free energy derivation from Luttinger-Ward functional}\label{app:b}
Following the prescription in Refs. \onlinecite{Luttinger1960,Rainer1976,Vorontsov2010}, we note that \bea\label{eq:app3}
F&=&-\int_K{\rm Tr}\left[\ln\left\{-G^{-1}_K\right\}\right]-\frac12\int_K{\rm Tr}\left[\Sigma_K G_K\right]\nonumber\\
&\equiv&~~~F_1+F_2,
\eea
where $G_K$ is the Greens' function given by $G^{-1}_K=[G^{0}_K]^{-1}-\Sigma_K$, and
\beq\label{eq:Green}
G^0_K=\left(\begin{array}{cc}
\frac{1}{i\omega_n-\e_{\bk}}&0\\
0&\frac{1}{i\omega_n+\e_{\bk}}
\end{array}\right),
\eeq
with $\omega_n\rightarrow
(2n+1)\pi$ and the self energy $\Sigma_K$ is 
\beq\label{eq:Self}
\Sigma_K=\left(\begin{array}{cc}
\bar\Phi_0f_{\bk}&-\Delta_0\mathcal{Y}_{\bk}\\
-\Delta^*_0\mathcal{Y}_{\bk}&-\bar\Phi_0f_{\bk}
\end{array}\right).
\eeq
Thus, 
\beq\label{eq:GreenF}
G_K=\frac{1}{\omega_n^2+E^2_{\bk}}\left(\begin{array}{cc}
-(i\omega_n+\bar\e_{\bk})&\Delta_0\mathcal{Y}_{\bk}\\
\Delta^*_0\mathcal{Y}_{\bk}&-(i\omega_n-\bar\e_{\bk})
\end{array}\right).
\eeq

The term $F_1$ can be computed as
\bea\label{eq:pot} F_1 &=& -\int_K{\rm
Tr}\left[{\rm ln}\left\{-G^{-1}\right\}\right]\nonumber\\
&=&-\int_K{\rm Tr}\left[{\rm
ln}\left\{\mathcal{H}-i\omega_n\right\}\right]\nonumber\\
&=& \int_K\int_{-i\omega_n}^{\infty} d\lambda{\rm
Tr}\left[\left\{\mathcal{H}+\lambda\right\}^{-1}\right]\nonumber\\
\Rightarrow\Delta F_1&=&\int_K\int_{-i\omega_n}^{\infty}
d\lambda{\rm Tr}\left[\left\{\mathcal{H}_{\rm
sc}+\lambda\right\}^{-1}-\left\{\mathcal{H}_{\rm
n}+\lambda\right\}^{-1}\right]\nonumber\\
&=&\int_K\int_{-i\omega_n}^{\infty} d\lambda
\left\{\frac{1}{\lambda-E}+\frac{1}{\lambda+E}-\frac{1}{\lambda+\e}\right.\nonumber\\
&&\left.~~~~~~~~~~~~~~~~~~~~~~~~~~~~~~~~~~~~~-\frac{1}{\lambda-\e}\right\}\nonumber\\
&=&-\int_K\left[{\rm ln}(-i\omega_n-E/T)+{\rm
ln}(-i\omega_n+E/T)\right.\nonumber\\
&&~~~~~~~\left. -{\rm ln}(-i\omega_n+\e/T)-{\rm ln}(-i\omega_n-\e/T)\right]\nonumber\\
&=&- T\int_{\bk}{\rm ln}
\left[\frac{(1+e^{E/T})(1+e^{-E/T})}{(1+e^{\e/T})(1+e^{-\e/T})}\right]\nonumber\\
&=&- T\int_{\bk}{\rm ln}
\left[\frac{\cosh^2(E/2T)}{\cosh^2(\e/2T)}\right].
\eea
Similarly the second part of the free energy yields
\bea
\Delta F_2&=&
\int_{\bk}\frac{\Delta_0^2\mathcal{Y}^2_{\bk} + \bar\Phi_0f_{\bk}\bar\e_{\bk}}{2E_{\bk}}\tanh\frac{E_{\bk}}{2T},\eea
where we have used that
\bea &&\sum_n e^{i\omega_n\eta^+}{\rm
ln}\left[i\omega_n-A\right]\nonumber\\
&=&-\int_C\frac{dz}{2\pi
i}e^{z\eta^+}n_{F}(z){\rm
ln}\left[z-A\right]\label{eq:com1}\nonumber\\
&=& -\left\{\int_{-\infty+i\delta}^{A+i\delta}\frac{dz}{2\pi
i}n_{F}(z){\rm ln}\left[z-A\right]\right.\nonumber\\
&&~~~~~~~+\left.\int^{-\infty-i\delta}_{A-i\delta}\frac{dz}{2\pi i}n_{F}(z){\rm
ln}\left[z-A\right]\right\}\label{eq:com2}\nonumber\\
&=& -\left\{\int_{-\infty}^{A}\frac{dz}{2\pi i}n_{F}(z){\rm
ln}\left[z-A + i\delta\right]\right.\nonumber\\
&&~~~~~~~-\left.\int_{-\infty}^{A}\frac{dz}{2\pi
i}n_{F}(z){\rm
ln}\left[z-A-i\delta\right]\right\}\label{eq:com3}\nonumber\\
&=&-\int_{-\infty}^{A}\frac{dz}{2\pi i}n_{F}(z)\left\{{\rm
ln}\left[z-A + i\delta\right]-{\rm ln}\left[z-A -
i\delta\right]\right\}\label{eq:com4}\nonumber\\
&=&-\int_{-\infty}^{A}dz~n_{F}(z)\nonumber\\
&=&{\rm ln}\left[1+e^{-A}\right]-{\rm ln}~\infty.\eea
 The apparently undefined $\ln\infty$ cancels out in all physical calculations when one calculates any free energy difference.
\subsection{Positive definiteness of the coefficient in Eq. \ref{eq:X5}}\label{app:c}
\begin{figure}[htp]
$\begin{array}{c}
\includegraphics[width=0.9\columnwidth]{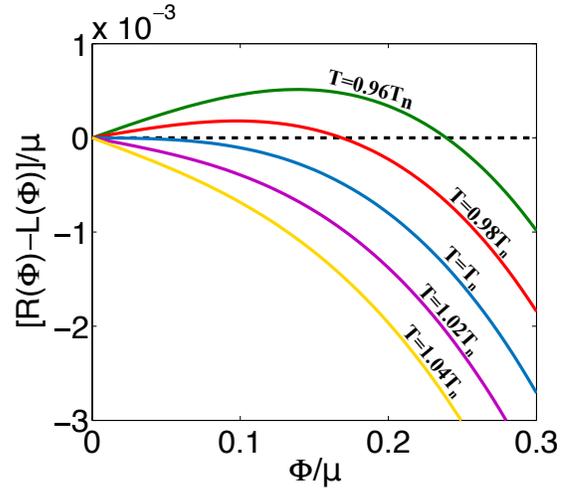}
\end{array}$\caption{
\label{fig:RPhi} Behavior of $R(\Phi)-L(\Phi)$ as function of $\Phi$ at different temperatures around $T_n$. Here we take a parabolic electron band in normal state as an example and used $\lambda_{\rm nem}=1.05$.}
\end{figure}

Here we show that the quantity
\beq\label{eq:Appe1}
Q=\frac{V^{\rm nem}}{2}\sum_{\bk}f^2_{\bk}\frac{{\rm sech}^2[\tilde\e_{\bk}/2T]}{2T},\nonumber
\eeq
which appears in Eq. \ref{eq:X5} of the main text,
is always less than unity. Note that this quantity does not know anything about the superconducting state, and $\Phi_0$ in $\tilde\e_{\bk}$ is the positive solution to
\beq\label{eq:Appe2}
\Phi=\frac{V^{\rm nem}}{2}\sum_{\bk}f_{\bk}\left[ \tanh\frac{\e_{\bk}+\Phi f_{\bk}}{2T}-1\right].
\eeq
Let us introduce $L(\Phi)\equiv $ LHS of Eq. (\ref{eq:Appe2}) $ = \Phi$ and $R(\Phi)\equiv$ right hand side of Eq. (\ref{eq:Appe2}). This equation has two non-negative solutions at any $T<T_n$: $\Phi=0$ and $\Phi=\Phi_0$. Differentiating $R(\Phi)$ with respect to $\Phi$ we get
\beq\label{eq:a3}
\frac{dR}{d\Phi}=\frac{V^{\rm nem}}{2}\sum_{\bk}f^2_{\bk}\frac{{\rm sech}^2[(\e_{\bk}+\Phi f_{\bk})/2T]}{2T}.
\eeq
Notice that $dR/d\Phi|_{\Phi=\Phi_0}=Q$. Treating $dR/d\Phi|_{\Phi=0}$ as a function of $T$, and taking $\e_{\bk}=k^2/2m-\mu$ as an example, this function is monotonically decreasing with increasing $T$. When $T {\ge} T_n$, the quantity $dR/d\Phi|_{\Phi=0}\le 1$, with the equality taking place at $T=T_n$. This can also be seen from Eq. (\ref{eq:MF2}). When $T<T_n$, $dR/d\Phi|_{\Phi=0}>1$, which means that $R(\Phi)$ starts above $L(\Phi)$ near $\Phi=0$ as shown in Fig. \ref{fig:RPhi}. What can also be proved is that for any $T$, $dR/d\Phi|_{\Phi\rightarrow+\infty}=(1/2)\lambda_{\rm nem}<1$ and $d^2R/d\Phi^2<0$ at any positive $\Phi$. This means that as $\Phi$ increases, $R(\Phi)$ crosses $L(\Phi)$ from above at $\Phi=\Phi_0$ for any $T<T_n$. This guarantees $dR/d\Phi|_{\Phi=\Phi_0}=Q<dL/d\Phi|_{\Phi=\Phi_0}=1$.

\end{document}